\documentclass[a4paper,twocolumn,superscriptaddress,groupedaddress,accepted=2025-09-29]{quantumarticle}
\pdfoutput=1
\usepackage[colorlinks,urlcolor=blue,linkcolor=blue]{hyperref}
\usepackage{graphicx}
\usepackage[dvipsnames]{xcolor}
\usepackage{amsmath}
\usepackage{dsfont}
\usepackage{amsfonts}
\usepackage{amssymb}
\usepackage{appendix}
\usepackage{cite}
\usepackage{xcolor}
\usepackage{soul}        
\setstcolor{red}         
\soulregister\cite7      

\newcommand{\pd}{{\phantom{\dagger}}}
\newcommand{\eps}{\varepsilon}

\newlength{\dhatheight}

\AtBeginDocument{%
  \pdfstringdefDisableCommands{%
    \def\cite#1{[#1]}%
    \def\eqref#1{(#1)}%
  }%
}

\hypersetup{citecolor=red}

\begin{document}

\title{Symmetry resolved out-of-time-order correlators of Heisenberg spin chains using projected matrix product operators}

\author{Martina Gisti}
\email{martina.gisti@uni-bonn.de}
\author{David J. Luitz}
\email{david.luitz@uni-bonn.de}
\author{Maxime Debertolis}
\email{maxime.debertolis@uni-bonn.de}
\affiliation{Institute of Physics, University of Bonn, Nu\ss allee 12, 53115 Bonn, Germany}

\begin{abstract} We extend the concept of operator charge in the context of an 
    abelian $U(1)$ symmetry and apply this framework to symmetry-preserving 
    matrix product operators (MPOs), enabling the description of operators 
    projected onto specific sectors of the corresponding symmetry. Leveraging 
    this representation, we study the effect of interactions on the scrambling 
    of information in an integrable Heisenberg spin chain, by controlling the 
    number of particles. Our focus lies on out-of-time order correlators (OTOCs) 
    which we project on sectors with a fixed number of particles. This allows us 
    to link the non-interacting system to the fully-interacting one by allowing 
    more and more particle to interact with each other, keeping the interaction 
    parameter fixed. 

While at short times, the OTOCs are almost not affected by interactions, the 
spreading of the information front becomes gradually faster and the OTOC 
saturate at larger values as the number of particle increases. We also study the 
behavior of finite-size systems by considering the OTOCs at times beyond the 
point where the front hits the boundary of the system. We find that in every 
sector with more than one particle, the OTOCs behave as if the local operator 
was rotated by a random unitary matrix, indicating that the presence of 
boundaries contributes to the maximal scrambling of local operators.
\end{abstract}

\maketitle

\section{Introduction}
\label{Introduction}
  The study of the dynamical process of information scrambling in quantum 
many-body systems with local interactions is fundamental to understand the 
mechanism of thermalization in closed quantum systems undergoing unitary 
dynamics. It has recently attracted significant attention due to its profound 
implications for quantum chaos, thermalization, and entanglement dynamics. The 
key phenomenon in this context is the formation of an emergent light-cone in the 
unitary dynamics of local operators~\cite{Lieb_Robinson_1972}, dictating the 
causal structure of information propagation. This has direct consequences for 
the area law of entanglement, the decay of correlations, and characteristic 
timescales of thermalization~\cite{Hastings_2007,Eisert_2010, 
Bravyi_2006,Luitz_Lev_2017,Kastner_2017}.

A well-established observable for studying information scrambling is the 
out-of-time-order correlator (OTOC)~\cite{Larkin_1969, 
Maldacena_Shenker_Stanford_2016}, which quantifies the time evolution of the 
commutator between a Heisenberg operator and a static local operator serving as 
a probe.  

Information scrambling has been investigated in systems with classical chaotic
counterparts~\cite{Garcia-Mata_Saraceno_Jalabert_Roncaglia_Wisniacki_2018,Chavez-Carlos_Lopez-del-Carpio_Bastarrachea-Magnani_Stransky_Lerma-Hernandez_Santos_Hirsch_2019,Lakshminarayan_2019,Maldacena_Shenker_Stanford_2016,Rammensee_Urbina_Richter_2018,Rozenbaum_Ganeshan_Galitski_2017},
where the notion of quantum chaos~\cite{Garcia-Mata_Jalabert_Wisniacki_2023,Xu_Scaffidi_Cao_2020,Dowling_Kos_Modi_2023} was developed. Very significant advances are due to analytical work
on random circuits~\cite{Nahum_Vijay_Haah_2018,Von_Keyserlingk_Rakovszky_Pollmann_Sondhi_2018,Rakovszky_Pollmann_Von_Keyserlingk_2018,Khemani_Vishwanath_Huse_2018,Chan_De_Luca_Chalker_2018},
where predictions about the typical behavior of OTOCs in chaotic systems with and without diffusive particle transport have been derived. Their properties in energy conserving systems
following the eigenstate thermalization hypothesis (ETH) have also been
investigated~\cite{Huang_Brandao_Zhang_2019,Balachandran_Benenti_Casati_Poletti_2021,Capizzi_Wang_Xu_Mazza_Poletti_2025}.

Furthermore, OTOCs were used to study systems in or close to a many-body 
localized phase~\cite{Chen_2016,Swingle_Bentsen_Schleier-Smith_Hayden_2016, 
Huang_Zhang_Chen_2017,Fan_2017,Luitz_Lev_2017,Slagle_Bi_You_Xu_2017,
Chen_Zhou_Huse_Fradkin_2017}, where interactions between particles and disorder 
compete. These studies generally have to deal with the exponential growth of the 
Hilbert space dimension, in addition to the need for disorder averaging and 
hence require sophisticated numerical techniques like 
typicality~\cite{Luitz_Lev_2017,Colmenarez_Luiz_2020_Lieb_Robinson} and matrix 
product operators~\cite{Bohrdt_Mendl_Endres_Knap_2017,HemeryPollmannLuitz_2019}.

The simplicity of interpretation of OTOCs makes them a versatile tool also 
useful for studying integrable systems, where they help to characterize the 
spreading of entanglement~\cite{Iyoda_2018} and to distinguish them from non-integrable 
systems~\cite{Fortes_GarciaMata_Jalabert_Wisniacki_2019,Garcia-Mata_Jalabert_Wisniacki_2023,Gopalakrishnan_Huse_Khemani_Vasseur_2018}. 
In these cases, OTOCs provide insight into the relationship between dynamical quantum scrambling and 
integrability~\cite{Fan_2017,Fortes_GarciaMata_Jalabert_Wisniacki_2019}. To 
avoid dealing with the exponential numerical cost in interacting integrable 
systems, much research was focused on non-interacting quantum 
systems~\cite{Lin_Motrunich_2018a,Lin_Motrunich_2018b,Riddell_Sorensen_2019, 
Riddell_Sorensen_2020,Bao_Zhang_2020,Xu_Swingle_2020,Byju_Lochan_Shankaranarayanan_2023,Riddell_Kirkby_ODell_Sorensen_2023}, 
where the complexity scales polynomially with system size and allowing for 
analytical results to support numerical simulations~\cite{Lin_Motrunich_2018a, 
Xu_Swingle_2020}. While some features of OTOCs in non-interacting 
integrable systems, such as early power-law growth~\cite{Lin_Motrunich_2018a,Colmenarez_Luiz_2020_Lieb_Robinson,Riddell_Sorensen_2019_rand_XX,Fortes_GarciaMata_Jalabert_Wisniacki_2019}, 
are shared with interacting integrable systems, others differ: in 
non-interacting systems, the late times saturation value of OTOCs goes to zero, 
whereas it stabilizes at a finite value in interacting systems. This 
raises the question of how these two limits are connected, which is the focus of 
this work. Here, we study OTOCs projected onto sectors with different fermionic 
particle numbers as a way to control the degree of interactions in the system, 
where both the non-interacting and fully-interacting cases are recovered in the 
respective limits of one particle and half-filling.

The computation of OTOCs for one-dimensional interacting clean systems has 
previously been performed using matrix product operators 
(MPOs)~\cite{Bohrdt_Mendl_Endres_Knap_2017,HemeryPollmannLuitz_2019,Xu_Swingle_2020,Lopez_Piqueres_Ware_Gopalakrishnan_Vasseur_2021}. 
The time evolution is usually performed by using the time-evolving-block 
decimation (TEBD) for nearest-neighbour 
interactions~\cite{Vidal_TEBD_2003,schollwock_density-matrix_2011}, leveraging 
the trivial structure of the operator outside the light-cone. Indeed, the 
spatial support of a local Heisenberg operator evolved to time $t$ is restricted 
to a spatial region $|x|<v_{B}t$ (with $v_{B}$ the light-cone front speed) with exponentially suppressed tails.

Hence, the operator remains in a tensor product form with the region outside of the light-cone to a good approximation and the tails can be captured by a small bond-dimension. Here, we
focus on systems with particle number conservation  in which the $U(1)$ symmetry leads to a block sparse structure of the tensors, reducing the computational cost and storage
requirements~\cite{Singh_Sukhwinder_2012,Paeckel_Koehler_Manmana_2017,Parker_Cao_Zaletel_2020, Singh_Pfeifer_Vidal_2011, Guo_Poletti_2019,
johannes_hauschild_frank_pollmann_efficient_2018, itensor}.  Exploiting this structure, OTOCs can be computed in an efficient way for large systems in the region outside the light-cone
with low bond dimensions. Once the information front is reached, the bond-dimension shows a steep increase and the late time regime in the bulk of the light-cone requires exponentially
large bond-dimensions in the size of the light-cone region.

In order to study the symmetry resolved OTOCs, we need to project the operator to a specific symmetry sector within the MPO formalism.  The symmetric MPO representation for abelian
symmetries encodes the block structure of operators commuting with the generator of the symmetry group and reduces the computational effort on the non-zero elements. In this work, we
introduce an adaptation of the traditional symmetric MPO formalism~\cite{Singh_Pfeifer_Vidal_2011, Guo_Poletti_2019, johannes_hauschild_frank_pollmann_efficient_2018, itensor} enforcing
additional constraints, which provides a direct representation of a \emph{single} block of an operator with a well defined operator charge as an MPO. These new constraints further
reduce the computational cost by highlighting extra simplifications in the internal block structure of the MPO, which are extremely efficient in the sectors with a few particles in
which the constraints are stronger. Besides the computational advantage, this representation permits the study of operators projected onto different symmetry sectors. This is required
for the calculation of the symmetry resolved OTOC which is at the center of our interest.

The paper is organized as follows: In Section~\ref{sec:Operator_charge}, we 
generalize the notion of operator charge, which provides a refined resolution of 
operators for different kind of block structures. In 
Section~\ref{sec:Projected_MPO}, we develop a set of constraints on the virtual 
spaces of MPOs adapted to a given definition of the operator charge, in order to 
construct MPOs projected onto a given symmetry sector. In 
Section~\ref{sec:OTOC}, we study the projected OTOCs of a clean Heisenberg spin 
chain to distinguish between few-body behavior analogous to dynamics in 
non-interacting systems, and many-body diffusive behavior resulting from 
repeated particle scattering. We investigate the behavior of projected OTOCs in 
different time regimes and discuss the effect of interaction in each case. 
Additionally, we anticipate the saturation of OTOCs at late times when the 
boundary conditions play an important role, by providing an analytical formula 
for an OTOC undergoing a random unitary evolution, suggesting a 
boundary-assisted thermalization of local operators.

\section{Operator charge}
\label{sec:Operator_charge}
\begin{figure*}[t]
	\includegraphics[width=1.\textwidth]{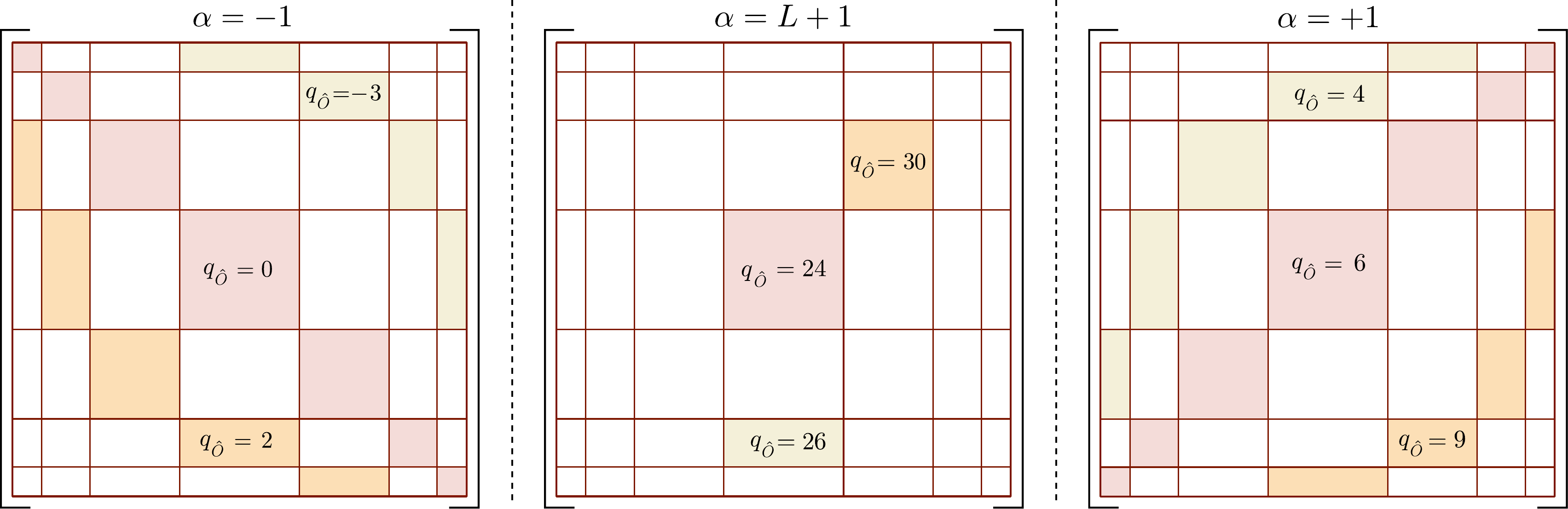}
	\caption{Sketch of the meaning of the generalized operator $\alpha$-charge defined by different values of $\alpha$, for $L=6$. Each parametrization corresponds to a
	different block structure defined by $\alpha$ and the corresponding $\alpha$-charge $q^{\alpha}_{O}$ of the operator $\hat{O}$, whose eigenvalues are indicated by a single color
	in each panel. \emph{Left}: $\alpha=-1$ corresponds to the standard definition of the operator charge. \emph{Middle}: $\alpha=L+1$ is the smallest integer value allowing a
	single-block $(n,n')$ resolution. \emph{Right}: $\alpha=+1$, allows to define blocks together on the anti-diagonal.}
	\label{fig:operator_charge}
\end{figure*}
	
Our work primarily focuses on operators with a block structure stemming from an 
abelian $U(1)$ symmetry, in such a way that their matrix representation contains 
non-zero elements only within identifiable blocks. Our discussion starts by 
recalling fundamental aspects of the representation of the abelian Lie group 
$U(1)$. The action of $U(1)$ on a vector space $\mathbb{V}$ naturally decomposes 
it into a direct sum of the group irreducible representations (irreps) 
$\mathbb{V} \cong \bigoplus_{n} \mathbb{V}^{\pd}_{n}$. Each $\mathbb{V}_n$ is a 
subspace (or \textit {sector}) of dimension $D^{\pd}_{n}$, associated to a 
charge $n \in \mathbb{Z}$.  Generators of $U(1)$ are elements of the 
corresponding algebra, $\hat{n} = \sum_n n ~ \hat{\Pi}_n$, where $\hat{\Pi}_n$ 
is the projector onto $\mathbb{V}_n$.  In quantum systems, $n$ defines the 
\textit{quantum charge} which is preserved by the symmetry.  For concreteness, 
we interpret $n$ as the  number of particles in the system which is conserved 
under the considered Hamiltonian time evolution.

In general, an operator $\hat{O}: \mathcal{H} \to  \mathcal{H}$ maps a state 
with a well defined charge $n$ into one with charge $n^{\prime}$, namely 
$\hat{O} = \sum_{i, j} \sum_{n,n^{\prime}} O^{(n',n)}_{i j} |i,  n^{\prime} 
\rangle \langle j , n | $. As an example, consider the creation operator 
$\hat{a}^{\dagger}$ adding a single particle to a state, and whose matrix 
representation exhibits a block structure where all blocks $O^{(n',n)}$ are 
empty unless $n' = n + 1$.  We can define a \textit{superoperator} $\mathcal{Q}$ 
as the commutator between the operator and the symmetry generator, i.e. 
$\mathcal{Q}[\cdot]=[\hat n,\cdot  ]$, such that:
\begin{equation}
	\begin{split}
	       {\mathcal{Q}} [\hat{a}^{\dagger}  ] & = \sum_{i, j, n}  {a}_{i j} ~ \bigg(\hat{n} | i,  n+1 \rangle \langle j , n   | -  | i,  n+1 \rangle \langle j , n    |
	       \hat{n} \bigg)  \\ &= (n + 1 - n) ~ \hat{a}^{\dagger}  = + \hat{a}^{\dagger} ,
	  \label{eq:superoperator_charge_action}
	\end{split}
\end{equation}
where $a_{i,j}$ are the matrix elements of $\hat{a}^{\dagger}$.  The operator $\hat{a}^{\dagger}$ can be interpreted as an \emph{eigenoperator} of $\mathcal Q$ with eigenvalue
$+1$ (the operator's charge), with the effect of creating \emph{one} single particle.  In the same manner, an operator $\hat O$ creating $q$ particles carries a charge $q$:
\begin{equation}
  \mathcal{Q}[\hat O] = \left[\hat n, \hat O\right]  = q \hat O.
  \label{eq:superoperator_eigenvalue}
\end{equation}
We now generalize this framework to include a broader class of operators that do not necessarily create or annihilate a specific number of particles (i.e., those with non-zero
blocks where $n'-n$ assumes integer values), yet still exhibit a recognizable block structure in the eigenbasis of $\hat{Q}$. Given two generic operators $\hat{O}_1$ and
$\hat{O}_2$, we introduce, through a free integer parameter $\alpha$, the generalized \textit{$\alpha$-commutator}:
\begin{equation}
	[\hat{O}_1 , \hat{O}_2 ]_{\alpha} = \hat{O}_1 \hat{O}_2 +  \alpha ~ \hat{O}_2 \hat{O}_1.
\end{equation}
The generalized \textit{supercharge} ${\mathcal{Q}}^{}_{\alpha}$ is the superoperator for which $\hat{O} $ is an eigenvector with eigenvalue $q^{}_{\alpha}$:
\begin{equation}
	\mathcal{Q}^{}_{\alpha}[\hat O] = \left[\hat n, \hat O\right]_{\alpha}  = q_{\alpha} \hat O.
	\label{eq:gen_op_charge}
\end{equation}
Then, without losing generality, the action of the generalized $\mathcal{Q}^{}_{\alpha}$ on $\hat O$ is the sum of a charge operator acting on the columns and the charge
operator acting on the rows, 
\begin{equation}
	\begin{split}
	       {\mathcal{Q}}^{}_{\alpha} [\hat O] &=\sum_{i, j, n, n'}  O^{[n',n]}_{i j} \bigg( \hat{n} | i , n' \rangle \langle j , n  | + \alpha 
		| i , n' \rangle \langle j , n | \hat{n} \bigg) \\ 
	       &= (n^{\prime} + \alpha ~ n) ~ \hat{O}  = q^{\pd}_{\alpha} \hat{O},
	  \label{eq:superoperator_charge_action2}
	\end{split}
\end{equation}
generalizing the \textit{operator charge} to $q^{}_{\alpha} = n^{\prime} + \alpha ~ n$.

The role of the parameter $\alpha$ is not accidental, but identifies different groups of blocks of the matrix representing the operator.  In fact, the conventional prescription
$\alpha=-1$ implies $q^{\pd}_{\alpha} = n^{\prime} -n$, and the charge $q^{\pd}_{\alpha}=0$ corresponds to an operator preserving the charge, being characterized by non-zero
blocks along the main diagonal.  On the other hand, operators such as the generator of the particle-hole symmetry have non-zero blocks on the main anti-diagonal and would be 
described by $\alpha = +1$ (i.e., $q^{\pd}_{\alpha} = n^{\prime} + n$), and labelled with $q^{\pd}_{\alpha} = L$. 
If an operator is only non-zero in a single block, $\alpha=(L+1)$ is considered to assign unique charge $q^{\pd}_{\alpha}=n^{\prime} + (L+1) n$ to each block. 
When this block lies on the main diagonal, its charge is simplified to $q^{\pd}_{\alpha} = n(L+2)$. In Fig.~\ref{fig:operator_charge}, we illustrate how different block 
structures of an operator $\hat{O}$ are associated to different values of the parameter $\alpha$.
Although the notion introduced here for generalized operator charge is flexible, not every choice of $\alpha$ admits a clear physical interpretation, and its value should be 
determined depending on the chosen operator. The proposed linear parametrization can be extended to any function $f(n,n')$ in order to capture more elaborate block structures
of operators.

In the following section, we will refer to an operator $\hat{O}$ as \textit{charge-preserving} if the parametrization of the supercharge is $\alpha=-1$ and $q^{}_{\alpha}=0$.  and
consequently that can be decomposed as:
\begin{equation}
	\hat{O} = \bigoplus_{n} \hat{\Pi}_n \hat{O} \hat{\Pi}_n = \bigoplus_{n} \hat{O}^{[n]},
	\label{eq::ShurLemma}
\end{equation}
meaning that the action of a charge-preserving operator on a state with a well-defined charge $n$ is reduced to the action of $\hat{O}^{[n]}=\hat{\Pi}_n \hat{O} \hat{\Pi}_n$ on
the corresponding subspace $\mathbb{V}_n$.  While $\hat{O}$ is associated to $\alpha=-1$ and $q^{}_{\alpha}=0$, the projected operator $\hat{O}^{[n]}$ requires the
parametrization $\alpha=L+1$ and the charge $q^{}_{\alpha}=n(L+2)$. In the following section, we focus on the MPO representation of these projected operators according to the
definition of operator charge in Eq.~\eqref{eq:gen_op_charge}.

\section{Projected MPO}
	\label{sec:Projected_MPO}
	In this section, we introduce the previously defined generalized operator charge to the matrix product operator (MPO) formalism.  Although this discussion focuses on
	one-dimensional MPOs, the theoretical framework can be generalized to higher dimensions. We review the implementation of abelian $U(1)$ symmetries in local tensors and adapt it
	to preserve the generalized operator charge throughout symmetry-preserving dynamics.  Charge-preserving MPOs are constructed by imposing constraints on the virtual indices,
	ensuring that the state it operates on maintains its global charge~\cite{Singh_Sukhwinder_2012,johannes_hauschild_frank_pollmann_efficient_2018}.  We will henceforth utilize the
	fundamental properties of MPOs and adopt the standard graphical notation~\cite{schollwock_density-matrix_2011,Crosswhite_Bacon_2008,perez_matrix_2007,verstraete_matrix_2008}.
	
\begin{figure}
                \includegraphics[width=1.\columnwidth]{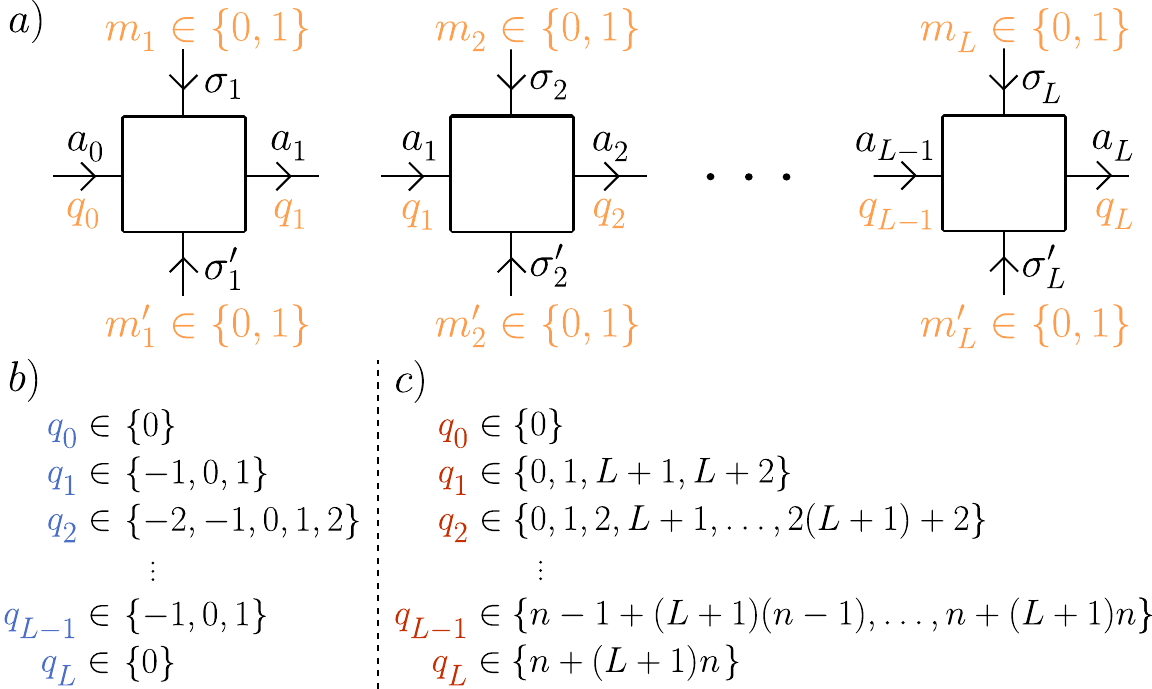}
		\caption{a) Diagrammatic representation of a symmetric MPO, in which each vector space is decorated by a charge.  $m$ denotes the charge of physical vector spaces, and
		$q$ the charge of virtual ones. 
		b) Charge-preserving MPOs ($\alpha=-1$):  allowed charges of virtual indices $a_i$, encoding the difference of accumulated local charges  $q^{\pd}_i = \sum^{i}_{j=1}
		m^{\prime}_j- m_j$.  
		c) Projected MPOs ($\alpha=L+1$): the corresponding charge becomes $q_i=\sum^{L}_{j=1} m^{\prime}_i+(L+1)m^{\pd}_i$.  The charge of the operator in the generalized
		definition is imposed at the boundary: $q^{}_L = n' + \alpha n$. } 
        \label{fig:MPO_representation}
	\end{figure}

	In general, an operator $\hat{O}$ acting on the Fock space of $L$ sites containing up to $L$ spinless fermions on a one-dimensional chain can be expressed as an MPO as follows:
	\begin{equation}
		\hat{O} = 
		\sum_{ \substack{\sigma_1,\dots,\sigma_L \\  \sigma_1^{\prime},\dots,\sigma_L^{\prime} \\ a_0, \dots,a_L } }  
		{O}^{\sigma_1, \sigma_1^{\prime} }_{a_0, a_1} \dots
		{O}^{\sigma_L, \sigma_L^{\prime} }_{a_{L-1}, a_L}
		\left|  \sigma_1 ,.., \sigma_L \rangle 
		\langle \sigma_1^{\prime},.., \sigma_L^{\prime} 
		\right|.
	\end{equation} 
	The above sums run over physical indices labelled by $\sigma$ and $\sigma'$ and virtual ones labelled $a^{\pd}_i$. The graphical illustration of the above expression is sketched
	in Fig.~\ref{fig:MPO_representation}~a) outlined in black.  The state $|\sigma_{i}\rangle$ corresponds to the ingoing physical index (or physical \textit{leg}) of the tensor at
	site $i$ and represents the states of the local direct Hilbert space $\mathcal{H}_{i}$ of the particle (or spin) at this site.  The state $\langle\sigma'_{i}|$ is related to the
	outgoing physical leg of tensor at site $i$, acting on the dual space of $\mathcal{H}_{i}$.  Hence, in the matrix representation of an operator, $|\sigma_{i}\rangle$ and
	$\langle\sigma'_{i}|$ stand for the row and column indices, respectively. The virtual leg $a_{i}$ connects the local site-$i$ tensor to the site-$(i+1)$ tensor and encodes
	information about the entanglement between left and right parts of the operator, split between sites $i$ and $i+1$. 

	To enforce the symmetry conditions, a charge value is associated to each vector space~\cite{Singh_Pfeifer_Vidal_2011,Singh_Sukhwinder_2012}.  We denote the sectors of physical
	indices by their charge $m_i$ (for spin-$1/2$, we consider $m_i\in\{0,1\}$ equivalently to $\{-1/2, 1/2\}$) and the state index by $\gamma^{\pd}_{m_i}$, with $\gamma^{\pd}_{m_i}
	\in [1,D_{m_{i}}]$. $D_{m_i}$ is the degeneracy of the eigenvalue $m_i$ of the charge operator, which counts for the maximal dimension of the corresponding vector space. For
	virtual indices, the charge and degeneracy are designated as $q_{i}$ and $d_{q_{i}}$. The indices are thus decorated as:
	\begin{equation}
		\sigma_i \rightarrow (m^{\pd}_i, ~\gamma^{\pd}_{m_i}), \quad a^{\pd}_i\rightarrow (q^{\pd}_i , ~ \gamma^{\pd}_{q_i}).
		\label{eq:loccharge}
	\end{equation}
	Building upon the idea of a charge-preserving MPS~\cite{Singh_Pfeifer_Vidal_2011}, we construct a charge-preserving MPO by keeping track of the flow of accumulated charges in
	the direct and dual spaces throughout the chain. In terms of the local quantum charges, we recall that a symmetric operator has a well-defined operator charge:
	\begin{equation}
		q^{\pd}_{\hat{O}} = \sum_{i=1}^{L}  m^{\prime}_i + \alpha ~ m^{\pd}_i ,
		\label{eq:constraints}
	\end{equation}
	with $ \sum_{i=1}^{L} m^{\pd}_i = n $ and $\sum_{i=1}^{L} m^{\prime}_i = n^{\prime}$. These sums can be split at any site $i$ to highlight the effect of the global charge at the
	level of local tensors: 
	\begin{equation}
		q_{i}^{\pd} = \sum_{j=1}^{i}m_{j}^{\prime} + \alpha ~ m^{\pd}_{j}.
		\label{eq:virtual_charge}
	\end{equation}
	Local charges are defined by the same parametrization of the supercharge $\alpha$, related to the fact that the reduced \emph{super density matrix} associated to the operator is
	also decomposed into blocks in eigenspaces of the supercharge operator split in two
	subsystems~\cite{Rath_Vitale_Murciano_Votto_Dubail_Kueng_Branciard_Calabrese_Vermersch_2023,Murciano_Dubail_Calabrese_2024}.  Equations for the charge flow at any site $i$
	follow naturally:
	\begin{equation}
		q^{\pd}_{i} = m^{\prime}_{i} + \alpha ~ m^{\pd}_{i} + q^{\pd}_{i-1}.
		\label{eq:charge_flow}
	\end{equation}
	The fist \textit{dummy} leg $q_{0}$ is considered $q_{0}=0$, as shown in Fig.~\ref{fig:MPO_representation}, since no charges are present in the left subsystem of the first
	tensor. The definition can be equally imposed from the right edge if one consider the charges adding up from right to left. In order to obtain a symmetry-preserving MPO the
	above condition implies that each local tensor must obey the following constraint:
	\begin{equation}
		O^{\sigma_i, \sigma_i^{\prime}}_{a_{i-1}, a_i} = 
		O^{m_i, m^{\prime}_i}_{q_{i-1}, q_i} \delta^{\pd}_{m^{\prime}_i + \alpha  m^{\pd}_i+q_{i-1}, q_i}.
	\end{equation}

	The accessible sectors with charge $q^{}_i$ at site $i$ are limited because, even though $q^{}_i$ is not explicitly presented as bounded in Eq.~\eqref{eq:charge_flow}, for
	spin-$1/2$ particles $m^{\pd}_{i} - m^{\prime}_{i}$ only takes integer values (we recall that $m_i\in\{0,1\}$), i.e. $m^{\pd}_{i} - m^{\prime}_{i} \in \lbrace
	-1,0,1\rbrace$.  In fact, using Eq.~\eqref{eq:charge_flow} iteratively from the left edge $q_0$ to the right and independently from $q_L=n'+\alpha n$ to the left, and then
	calculating the overlap of the two resulting sets of allowed sectors at each site $i$ makes it straightforward to identify the inaccessible symmetry sectors for an MPO.

	\begin{figure}[t]
		\includegraphics[width=1.\columnwidth]{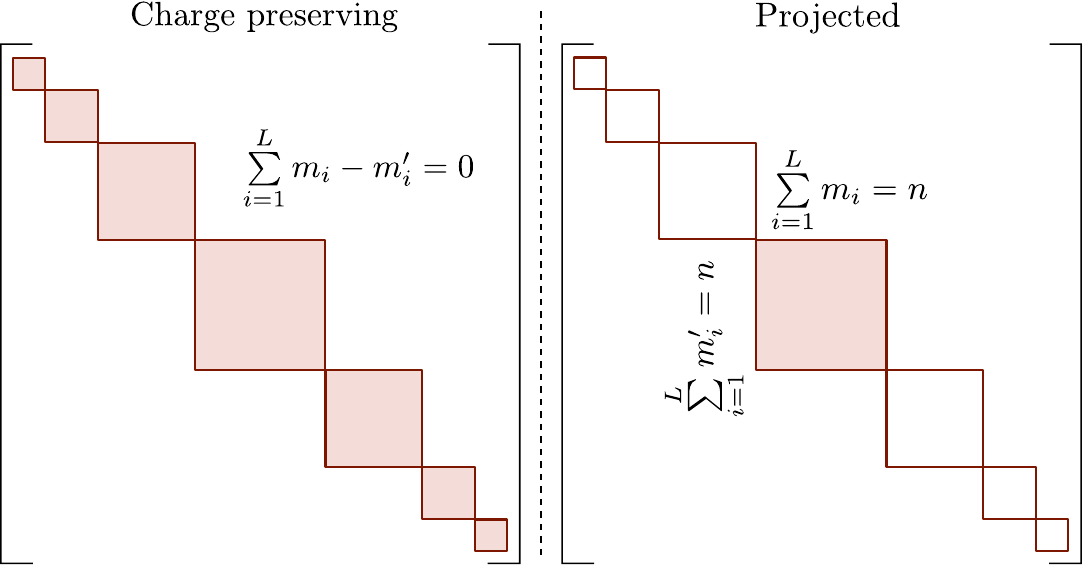}
		\caption{ \textit{Left}: Matrix representation of a charge-preserving operator. 
		Only the non-zero elements are highlighted, and every sector (in the main diagonal) is spanned. 
		\textit{Right}.  Projected representation of an operator, equivalent to the projection of the operator within a given symmetry sector. Every element outside the
		highlighted sector are projected out.} 
		\label{fig:sector_vs_projected}
	\end{figure}

	The purpose of this work is to investigate how quantum information spreads in sectors with a fixed number of particles by using sector projected MPOs. Hence, we adopt the
	prescription $\alpha=(L+1)$ in the following and concentrate on operators with a well-defined operator charge: $q^{}_{\hat{O}}=n(L+2)$. In theory, there are $(L+1)^{2}$ distinct
	internal charges $q^{}_{i}\in [\![ 0, (L+1)^{2}-1]\!]$ in a projected MPO, but, depending on the selected global charge sector $n$, many of these sectors are inaccessible due to
	the fixed charges at the edges: the first virtual leg is always chosen to hold a charge $q^{}_{0} = 0$, and the last one corresponds to the global operator charge
	$q^{}_{L}=n(L+2)$.  Meanwhile to express a charge-preserving operator ($\alpha=-1$), the first dummy index is required to be $q^{\pd}_{0} = 0$ and the last virtual leg to be
	$q^{\pd}_{L}=0$, so that the particle number flux of incoming state is the same as the particle number of the outgoing state.  In the latter case, the internal charges can
	explore $2L+1$ different charge sectors, namely $q_{i}\in[\![-L, L]\!]$.

	The conditions on local tensors stated above will lead to MPOs representing an operator that, in the overall Hilbert space picture, has non-zero elements in the selected
	sectors as dictated by the chosen operator charge. The left panel of Fig.~\ref{fig:sector_vs_projected} shows an operator with a block-diagonal structure when the conditions for
	a charge-preserving MPO are imposed ($\alpha=-1$), whereas the right panel of Fig.~\ref{fig:sector_vs_projected} depicts an operator with non-zero elements only a given sector
	$n$, with the conditions for the projected MPO ($\alpha=L+1$).
        
	It is relevant to notice that the projected MPOs on global sectors $n=0$ (and $n=L$) only have one possible sector at each site $i$, that is $q^{}_{i} = 0$ ($q^{}_{i} =
	i+(L+1)i$), such that the corresponding bond dimension remains $1$ along the chain, being consistent with the dimension of these sector ${D}_{n=0} = 1 $ (${D}_{n=L} = 1$).  The
	graphical notation for a charge-preserving MPO, with $\alpha=-1$, is sketched in the panel (a) of Fig.~\ref{fig:MPO_representation}.  Panel (b) exhibits the internal charge
	structure of the virtual legs as defined in Eq.~\eqref{eq:virtual_charge}.  In the panel (c) of Fig.~\ref{fig:MPO_representation}, we show the construction of the internal
	sectors for a projected MPO along the chain when $\alpha=(L+1)$. 

\section{Symmetry Resolved OTOC}
	\label{sec:OTOC}
	\subsection{Model and definitions}
	We consider the clean Heisenberg spin-$1/2$ chain of length $L$ with open boundary conditions, defined by the Hamiltonian:
	\begin{equation}
		\hat{H} = \sum_{i=1}^{L} \frac{J}{2}\left( \hat{\sigma}^{+}_{i} \hat{\sigma}^{-}_{i+1} + \hat{\sigma}^{-}_{i} \hat{\sigma}^{+}_{i+1}\right) +
		\Delta \hat{\sigma}^{z}_{i}\hat{\sigma}^{z}_{i+1},
		\label{eq:Heisenberg}
	\end{equation}
	where $\sigma^{\pm}_i = (\sigma^{x}_i \pm i \sigma^{y}_i)/2$ are the raising and lowering operators, and $\sigma^{z}_i$ the $z$-component of the Pauli matrices acting on site $i$.
	We consider in this paper the isotropic case $J=\Delta=1$ to maintain the competition between hopping along the chain and spin-polarization along the $z$-axis. This model is
	integrable and can be solved exactly using the Bethe ansatz~\cite{Giamarchi_Press_2004,Faddeev_1996}, and it constitutes for us a testbed for exploring properties of many-body
	systems emerging from interactions between spins (or equivalently particles). It is particularly appealing since simple analytical formulae can be derived in the one-particle
	sector (defined later), in which interactions are frozen, and confront them to the full many-body problem as the number of interacting degrees of freedom considered increases.

	This model is $U(1)$ symmetric, and the corresponding charge operator is defined as follows:
	\begin{equation}
		\hat{S}^{z}_{\mathrm{tot}}=\frac{1}{2}\sum_{i=1}^{L} \hat{\sigma}^{z}_{i}, \quad [\hat{S}^{z}_{\mathrm{tot}},\hat{H}]=0,
	\end{equation}
	such that the Hamiltonian is defined in blocks associated with the different eigenvalues of the charge operator related to the magnetization $m=\sum_{i=1}^{L} \langle
	\sigma^{z}_i \rangle = (n^{}_{\uparrow} - n^{}_{\downarrow})/2$, such that $m\in\{-L/2,...,L/2\}$. Alternatively, we label the sectors of $\hat{H}$ by the particle number
	$n=n^{}_{\uparrow}\in\{0,...,L\}$, which is relevant through the Jordan-Wigner transformation mapping spin degrees of freedom to fermionic ones. These two charges are related by
	$n=L/2+m$.
	
	In the following we are focusing on the $ZZ$-OTOC, defined as the squared commutator of two local operators evolved in time as:
	\begin{equation}
		\label{eq:def_otoc}
		\begin{split}
			C_{j, j^{\prime}}\left(t\right) &= \frac{1}{2} \left\langle \left[\hat{\sigma}^z_{j}(t), \hat{\sigma}^z_{j^{\prime}}(0) \right]^{2}\right\rangle\\ 
			&= 1 - \frac{\mathrm{Re}(\mathrm{Tr}\left[\hat{\sigma}^z_{j}(t) \hat{\sigma}^z_{j^{\prime}}(0) \hat{\sigma}^z_{j}(t)
			\hat{\sigma}^z_{j^{\prime}}(0)\right])}{\mathrm{Tr}[\mathds{1}]},
		\end{split}
	\end{equation}
	where $\mathrm{Tr}[\mathds{1}]=2^{L}$ is the dimension of the corresponding Hilbert space, and where we used the property of Pauli operators to be both hermitian and unitary. To
	study the operator for typical initial states, we consider the expectation value at infinite temperature, which results in taking the trace of the commutator. This observable
	quantifies the spatial spread (by varying $j'$) of the support of the initially local operator over time. In particular, we investigate the projected OTOC in a sector with $n$
	particles, $C^{[n]}_{j,j'}(t)$, for which each Pauli operator involved in Eq.~\eqref{eq:def_otoc} is represented by a projected MPO. The full OTOC is thus a combination of the
	projected ones:
	\begin{equation}
		C^{}_{j,j'}(t) = \sum_{n=0}^{L} \frac{D_n}{2^{L}_{}} \, C^{[n]}_{j,j'}(t),
		\label{eq:otoc_decomposition}
	\end{equation}
	where $D_n$ is the dimension of the subspace: 
	\begin{equation}
	D_n = \mathrm{Tr}[\mathds{1}^{[n]}_{}] = \frac{L!}{n!(L-n)!}.
		\label{eq:dimension_proj}
	\end{equation}
	Increasing the number of particles allows interactions to play a more important role in the commutator: we start by stating exact results in the one-particle sector for the OTOC
	and confront them later on to the interacting problem.
	
	\subsection{Exact results of the \texorpdfstring{$ZZ$}{ZZ}-OTOC with one particle}
	In the one-particle sector, the Hamiltonian is equivalent to a non-interacting fermionic tight-binding chain through the Jordan-Wigner transformation: 
	\begin{equation}
		\hat{H}^{[n=1]} = J \sum_{i=1}^{L} \left(\hat{c}^{\dagger}_{i} \hat{c}^{}_{i+1} + \hat{c}^{\dagger}_{i+1} \hat{c}^{}_{i}\right).
		\label{eq:free_H}
	\end{equation}
	The features of the OTOCs in free fermionic problems have been previously reported in various contexts~\cite{Lin_Motrunich_2018a,Lin_Motrunich_2018b,Riddell_Sorensen_2019,
	Riddell_Sorensen_2020,Bao_Zhang_2020,Xu_Swingle_2020,Byju_Lochan_Shankaranarayanan_2023,Riddell_Kirkby_ODell_Sorensen_2023}, and our discussion here primarily contextualizes
	these findings within our specific framework. These results serve as a benchmark for assessing deviations in many-body systems from the free-fermion case.
	Eq.~\eqref{eq:free_H} can be solved exactly, allowing us to derive the exact expression of the $ZZ$-OTOC defined in Eq.~\eqref{eq:def_otoc}, in the thermodynamic limit for
	a translation invariant system (see Appendix~\ref{appendix:free_model} for details):
	\begin{equation}
		C^{[n=1]}_{j,j'}(t) = 2^{5}\left(J_{l}^{2}(t) - J_{l}^{4}(t)\right).
		\label{eq:exact_OTOC}
	\end{equation}
We derive a simplified expression involving Bessel functions of the first kind 
$J_l(t)$, with the integer order $l=|j'-j|$ corresponding to the spatial 
separation between local operators, with time as argument. The OTOC thus 
exhibits oscillations both in space and time, along with an amplitude decay at 
late times time as a power law $J_{l}^{2}(t \rightarrow \infty) \propto t^{-1}$, as 
previously 
predicted~\cite{Lin_Motrunich_2018a,Lin_Motrunich_2018b,Bao_Zhang_2020,Xu_Swingle_2020}.  
The primary contribution in Eq.~\eqref{eq:exact_OTOC} comes from the squared 
Bessel function, resembling the Fraunhofer diffraction pattern of a single slit 
with oscillations occurring in time instead of space. As the information 
spreads and reaches site $l$, the time-dependent oscillating pattern of the OTOC 
emerges, driven by $J_{l}^{2}(t)$, akin to a diffraction process of the operator 
$S^{z}_{j+l}$ by the light-cone of $S^{z}_{j}(t)$. 

	\begin{figure}
		\includegraphics[width=1.0\columnwidth]{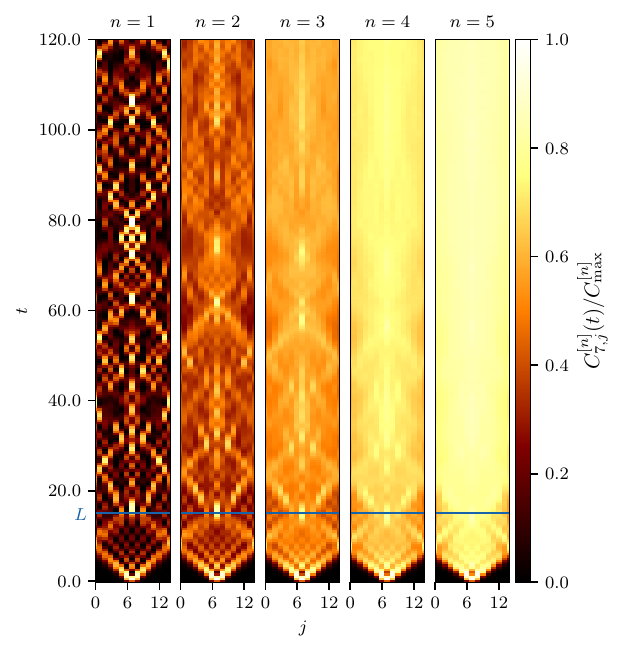}
		\caption{OTOC in different sectors for $L=15$, $dt=0.05$, without truncation. The OTOC are computed up to times far beyond the boundary reflection, in order to identify
		the different time regions. The normalization $C_{\mathrm{max}}^{[n]}$ allows the scale in each subplot to be similar, the amplitude being larger in bigger sectors. The
		horizontal blue line indicates the time $t=L$, at which the reflection hits back the site in the middle.}
		\label{fig:full_otocs}
	\end{figure}

\begin{figure*}[t]
        \includegraphics[width=1.\textwidth]{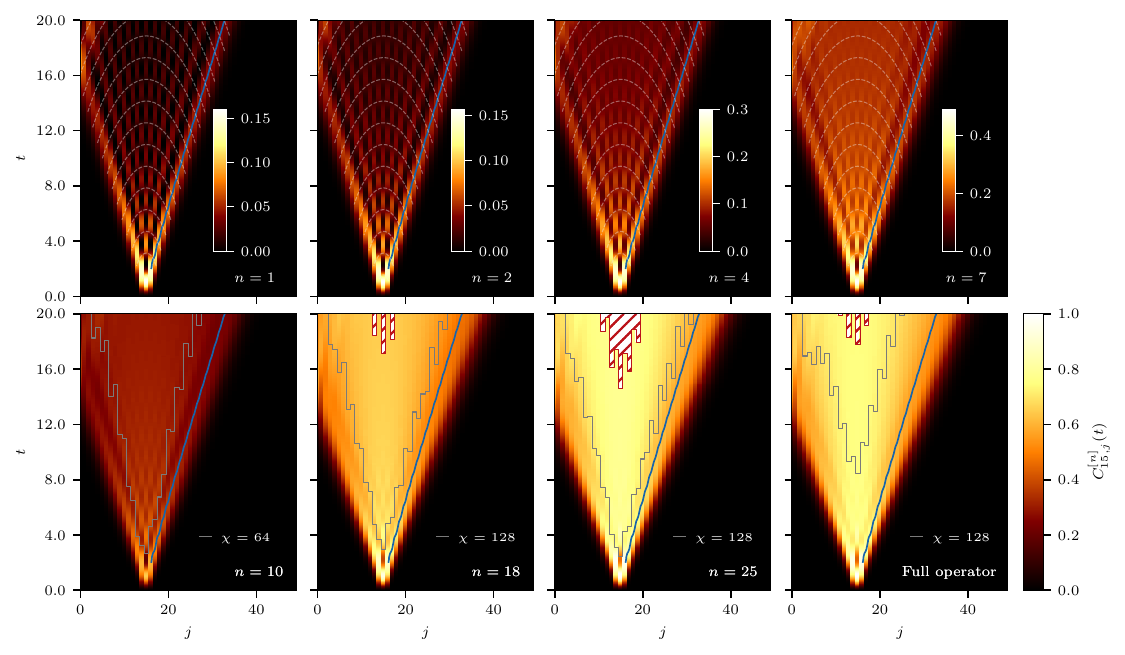} 
	\caption{OTOC of the isotropic Heisenberg chain projected in sectors with different number of particles $n$ for $L=50$, $\chi=6000$ and $\mathrm{d}t=0.005$ up to time $t=20$. The
	top-left panel shows the analytical result derived for free fermions (sector $n=1$). The $6$ following panels are time-evolved projected MPOs, for $n=2,4,7,10,18,25$, where the
	last one is the half-filled sector. The bottom-right panel corresponds to the full MPO representation (all sectors are considered) with operator charge $\alpha=-1$. The blue
	line in each panel indicates the front of the OTOC for $n=1$. In the first row, dashed lines indicate diffraction events in space-time in the $n=1$ sector.
	In the second row, red hatched regions inside the light-cone indicate that the cumulative truncated singular values squared (discarded weight) $\lambda^2$ at the corresponding
	bond of the MPO exceeds $10^{-4}$, and the simulations are thus not reliable there (computational light-cone). Grey lines indicate the position of the white hatched regions for
	simulations with a smaller bond dimensions ($\chi=64$ and $\chi=128$) as specified in the legends.}
	\label{fig:2D_otocs}
\end{figure*}

	The non-interacting OTOC is plotted in the top-left panel of Fig.~\ref{fig:2D_otocs}, where the function in Eq.~\eqref{eq:exact_OTOC} is plotted in space and time, revealing
	distinct constructive and destructive interference patterns. To compare the propagation in time with that in space, we include dashed lines defined by the following
	parametrization: $(x,y) = (j+t \cos(\theta), t \sin(\theta))$. This illustrates that information spreads faster as we are closer to the light-cone by a factor of $\sin(\theta)$.

	From Eq.~\eqref{eq:exact_OTOC}, we can extract the early time growth of the OTOC, which is known to be a power-law by expanding the Baker-Campbell-Hausdorff formula close to
	$t=0$~\cite{Smith_Knolle_Moessner_Kovrizhin_2019,Lin_Motrunich_2018a,Colmenarez_Luiz_2020_Lieb_Robinson,Riddell_Sorensen_2019_rand_XX,Fortes_GarciaMata_Jalabert_Wisniacki_2019}. In
	Appendix~\ref{appendix:early_times}, we retrieve the expected power law by deriving the exact form of the OTOC in the limit $t\rightarrow 0$:
	\begin{equation}
		\partial_t \log(C^{}_{j,j^{\prime}}) \approx \frac{\alpha_l}{t}, \quad \alpha_l \approx 2l\left(1-\frac{t^{2}}{4l(l-1)}\right),
		\label{eq:free_power_law}
	\end{equation} 
	showing that the exponent $\alpha_l$ gets slightly smaller than the expected $2l$ value at short but non zero times. 

	\subsection{OTOC in symmetry sectors}

	We discuss now the behavior of the $ZZ$-OTOC projected on symmetry sectors, which we investigate in different time domains. The OTOC is known to exhibit different behaviors at
	early times and late times~\cite{Garcia-Mata_Saraceno_Jalabert_Roncaglia_Wisniacki_2018,Fortes_GarciaMata_Jalabert_Wisniacki_2019,Lin_Motrunich_2018a,Lin_Motrunich_2018b}, with
	specificities depending on the system under consideration. 
	At early times, scrambling of local information leads either to an exponential growth of the OTOC~\cite{Garcia-Mata_Saraceno_Jalabert_Roncaglia_Wisniacki_2018,Chavez-Carlos_Lopez-del-Carpio_Bastarrachea-Magnani_Stransky_Lerma-Hernandez_Santos_Hirsch_2019,Lakshminarayan_2019,Maldacena_Shenker_Stanford_2016,Rammensee_Urbina_Richter_2018,Rozenbaum_Ganeshan_Galitski_2017},
	or to a non-universal power-law~\cite{Dora_Moessner_2017,Fortes_GarciaMata_Jalabert_Wisniacki_2019}. The presence of an exponential growth is dictated by the underlying Lyapunov 
	spectrum of the quantum dynamics and depends on whether the scrambling is governed by a single characteristic timescale~\cite{Hallam_Morley_Green_2019,Leontica_Green_2025}.
	At late times, OTOCs in quantum systems typically saturate to a finite value around which system-dependent aperiodic oscillations are observed.  This late time regime is an 
	indicator of the degree of integrability breaking~\cite{Fortes_GarciaMata_Jalabert_Wisniacki_2019, Garcia-Mata_Saraceno_Jalabert_Roncaglia_Wisniacki_2018}, and
	is studied for times before which the information front reaches the boundaries, such that finite size effect do not play a role: in the following, we refer to this regime as the
	thermodynamic limit or intermediate times, where the limit $L\rightarrow \infty$ is considered before $t\rightarrow \infty$. To this standard discussion, we append an extra time
	regime which accounts for the late-time properties of OTOC for a \emph{finite system}, where unlike the thermodynamic limit, $t\rightarrow \infty$ is taken before $L\rightarrow
	\infty$. We dub this time regime the finite size regime, or late times, in which revivals of the entanglement entropy of local operators have been observed in previous
	studies~\cite{Modak_Alba_Calabrese_2020,Alba_2024}. 

	\begin{figure}
		\includegraphics[width=1.\columnwidth]{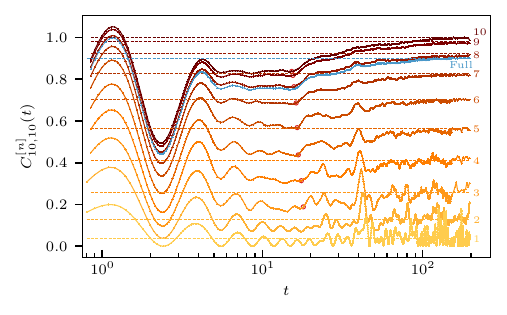}
		\caption{Time dependence of the OTOC projected on different sectors for $L=20$ obtained with typical states evolution (see text). The log scale allows to dissociate time
		regions, and red circles are indicators for the reflection on the boundary hitting the site in the middle. We indicate the number of particles corresponding to each curve by a
		number next to it with the corresponding color. The blue color labelled \emph{Full} corresponds to the full symmetric operator, defined as the sum over all sectors.}
		\label{fig:vertical_cuts}
	\end{figure}

	In Fig.~\ref{fig:full_otocs}, we present the OTOC in various sectors for $L=15$, calculated using projected MPOs without truncation. The results extend to times well beyond the
	point where the front of the OTOC has reached the open boundaries of the spin chain. In each sector, we can observe the intermediate and late time regimes: at times $t\lesssim
	L/2$, the OTOCs are not sensitive to the boundaries, until the end of the chain is hit by front of the light-cone (initially starting at site $j=L/2$), which \emph{bounces} back
	to its original location at time $t\simeq L$ with a slight dependence on the number of particles. From this time on, information keeps bouncing on the boundaries, leading to a
	second scrambling mechanism of information within the chain. The impact of particle number becomes significant in Fig.~\ref{fig:full_otocs}, where the distinct oscillations or
	diffraction patterns in the $n = 1$ sector gradually smooth out as $n$ increases, eventually disappearing altogether. This results in a uniform OTOC structure in both space
	and time.  

\begin{figure}
	\includegraphics[width=1.\columnwidth]{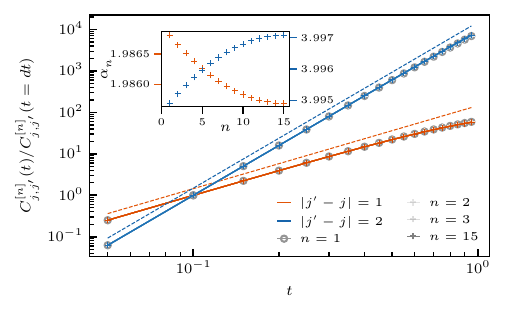}
	\caption{Early time OTOC ($t<1.$) for different sectors $n$ and different spatial separations $|j^{\prime}-j|$. Colored dashed lines show the predicted power law in the $n=1$
	sector.  \emph{Inset}: Fit of the power law in each sector, using  $C_{j,j'}^{[n]}(0\leq t \leq 0.2)$.}
	\label{fig:early_times_otocs}
\end{figure}

In order to study features of intermediate and late time regimes, we plot a 
vertical cut of the projected OTOC at $j'=j$ in 
Fig.~\ref{fig:vertical_cuts}, for a chain of $L=20$ sites where the 
intermediate regime extends longer. The time axis is plotted in logarithmic scale to 
squeeze the transient region between the first effect of the boundaries 
indicated by the red circles and the late time saturation. The simulation up to 
late times is computationally expensive for the projected MPO method, since the 
operator entanglement entropy of the time-evolved local operator grows linearly in 
time~\cite{Xu_Swingle_2020,Murciano_Dubail_Calabrese_2024}. Thus, we use the typical states evolution 
method~\cite{Luitz_Lev_2017,Colmenarez_Luiz_2020_Lieb_Robinson} to reach late 
times for system sizes until $L\simeq 30$, which does not rely on the 
low-entanglement structure of the operator. The projected MPOs are used to 
extract the early and intermediate times for larger systems, up to $L=50$ with a 
large bond dimension $\chi=6000$ to avoid truncation 
artifacts~\cite{Lopez_Piqueres_Ware_Gopalakrishnan_Vasseur_2021}.

In Fig.~\ref{fig:vertical_cuts}, we can see that after the initial scrambling 
process at early times, the OTOC displays an oscillatory pattern that gradually 
dampens over time and with increasing particle number $n$, settling around a 
fixed value that depends on $n$. Eventually, the front reflected at the 
boundary returns to site $j$, contributing further to scrambling, which induces a 
higher saturation value of the OTOC in the late-time regime. This  
sector-dependent saturation agrees well with the prediction from random matrix 
theory (dashed lines), which we discuss further on. This suggests complete 
scrambling of the corresponding operators in a finite 
system~\cite{Modak_Alba_Calabrese_2020,Alba_2024}, even as the system remains 
integrable. We now turn our attention to the effect of particle number across 
the three aforementioned time regimes.

\subsubsection{Early times}

In the early-time regime, the OTOC exhibits power-law growth. In Fig.~\ref{fig:early_times_otocs}, we present the OTOC projected onto different sectors with $n$ particles and spatial
separations of the evolving and probing operators $|j^{\prime}-j|$, showing very good agreement with Eq.~\eqref{eq:free_power_law}.  Notably, the early-time growth is dominated by the
behavior of the free problem ($n=1$), with minor deviations for sectors with more particles. From the Baker-Campbell-Hausdorff
formula~\cite{Colmenarez_Luiz_2020_Lieb_Robinson,Riddell_Sorensen_2019_rand_XX,Fortes_GarciaMata_Jalabert_Wisniacki_2019}, the exponent is expected to be the same in every sector for
$t\rightarrow 0$. In the inset of Fig.~\ref{fig:early_times_otocs}, we plot the fitted exponent $\alpha_{n}$ of the power low $t^{\alpha_n}$ for $t\in[dt,0.2]$ in each sector for both
separations $|j'-j|=1$ and $2$ to show that while the exponent is sensitive to interactions for $t>0$, the deviations from the $n=1$ sector remain small.  In
Appendix~\ref{appendix:early_times}, we conduct a finite-size scaling analysis of these deviations, demonstrating that they persist in the thermodynamic limit, despite the expected
independence of early-time dynamics from system size. Previous studies~\cite{Fortes_GarciaMata_Jalabert_Wisniacki_2019,Riddell_Sorensen_2019_rand_XX} have also found that early-time
behavior remains robust against transitions from integrable to chaotic behavior. 

\subsubsection{Intermediate times (thermodynamic limit)}
\begin{figure}
	\includegraphics[width=1.\columnwidth]{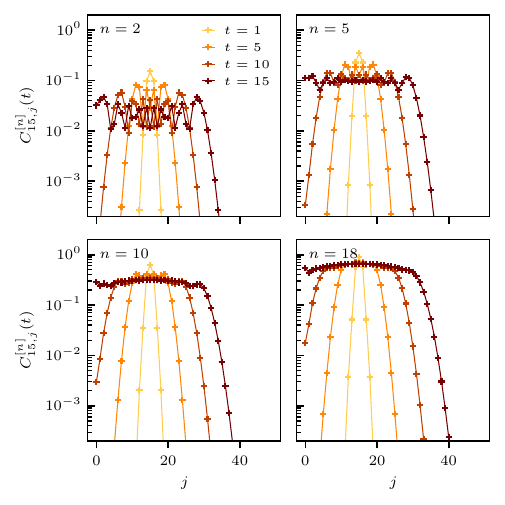}
	\caption{Spatial structure of the OTOC at different times and projections, for the same parameters as Fig.~\ref{fig:2D_otocs}.}
	\label{fig:horizontal_cuts}
\end{figure}
After the power-law growth at early times, the OTOC exhibits a space-time region 
dubbed ``light-cone'' inside which the commutator between the two operators 
$\hat \sigma^z_j(t)$ and $\hat \sigma^z_j$ is expected to be larger. We 
investigate here the effect of particle number on the properties of the front 
and on the saturation value of the OTOC deep inside the light-cone, in the 
thermodynamic limit. In Fig.~\ref{fig:2D_otocs}, we plot the spatio-temporal 
behavior of the OTOC for a chain of $L=50$ sites up to time $t=20$, with 
different particle number ranging from $n=1$ (free problem) to $n=25$ 
(half-filled sector). For the top panels and the three left-most panels in the 
bottom row, simulations correspond to projected MPOs (operator charge 
parametrized by $\alpha=L+1$), such that the trace in Eq.~\eqref{eq:def_otoc} 
runs over the corresponding sector. The bottom right-most panel corresponds to 
the full symmetric representation of the MPO ($\alpha=-1$), the trace being 
hence taken over the full $2^{L}$-dimensional Hilbert space. The same 
implementation of symmetric MPOs is used for both full and projected MPO 
simulations for comparison, where internal constraints vary according to the 
corresponding definition of the operator charge (see 
Eq.~\eqref{eq:charge_flow}). The bond dimension for projected MPO simulations is 
$\chi=6000$  and $\chi=1024$ for the full symmetric one. The convergence of the 
OTOC is tracked through the cumulative sum of truncated singular values 
$\varepsilon_{\mathrm{sum},i}$ at each bond $i$ over time. This allows us to 
monitor the convergence of MPOs in a given sector up to time $t$ for small 
systems, for which we perform benchmarking ED simulations. We use a convergence 
threshold of $\varepsilon_{\mathrm{sum,i}}=10^{-4}$, which is indicated by  
 white hatched regions in Fig.~\ref{fig:2D_otocs}. We also plot the convergence 
for smaller $\chi$, which gives converged results up to times short after the 
front~\cite{Xu_Swingle_2020} (see Appendix~\ref{appendix:benchmark} for a 
discussion). OTOCs are computed efficiently by using the locality of the 
operator acting on site $j'$ which is not evolved, and use the fact that trace 
of the product of MPOs $S^{z}_{j}(t)S^{z}_{j'}S^{z}_{j}(t)S^{z}_{j'}$ is 
proportional to identities for $j \neq j'$ by definition of the canonical 
form~\cite{Vidal_TEBD_2003,schollwock_density-matrix_2011}.

	\emph{Front:}
We observe in Fig.~\ref{fig:2D_otocs} that the clear fringes present in the $n=1$, the zeros of the Bessel function in Eq.~\eqref{eq:exact_OTOC}, are spreading faster  both in time and
space as the number of particle increases, until they become indistinguishable for the larger $n$. Note that the front in the non-interacting case also broadens in time as
$t^{1/3}$~\cite{Hunyadi_Racz_Sasvari_2004,Platini_Karevski_2005,Fagotti_2017,Collura_De_Luca_Viti_2018}.  The blue line plotted in each panel represents the front of the OTOC in the
$n=1$ sector and allows to discern that the front spreads faster with increasing particle number $n$ as the effect of interactions gets stronger. As the fringes get enlarged, their
contributions add up and the amplitude inside the light-cone increases, such that the front gets wider and thus more difficult to identify for large $n$ when time increases, as seen in
Fig.~\ref{fig:horizontal_cuts}. We observe that the front has a faster velocity for large $n$ than for $n=1$. This indicates that there is a qualitative difference in the front dynamics
in the interacting case compared to the free system in line with the findings in Ref~\cite{Lopez_Piqueres_Ware_Gopalakrishnan_Vasseur_2021}. 

\begin{figure}
	\includegraphics[width=1.\columnwidth]{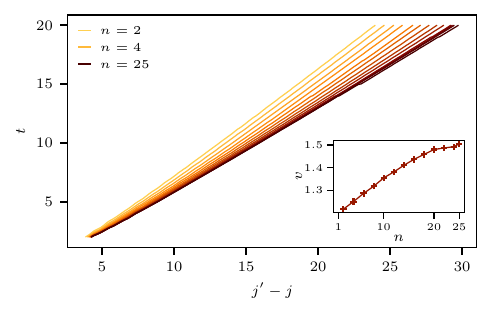}
	\caption{Extracted front propagation (refer to the text for a definition) for different projections for the same parameters as Fig.~\ref{fig:2D_otocs}. At each time step, the front
	is first spatial index whose value is below 1 percent of the maximum of the OTOC at this time step. \emph{Inset}: Front speed $v$ against charge sector $n$ extracted by
	a linear fit of $C^{[n]}_{j,j'}(t>16)$. } 
	\label{fig:speed}
\end{figure}

Thus, we compare the speed between sectors by following the point in space at which the OTOC reaches one percent of its maximum value at each time, as explained in
Appendix~\ref{appendix:front}. This value is plotted in the main panel of Fig.~\ref{fig:speed} for $L=50$ and different $n$. We choose $j=15$ and not $L/2$ in this case to see the front
spreading further in space in one direction before reaching the boundary. The OTOC spreads faster in space with larger $n$, in a non-linear fashion in all sectors for the time
considered here. Indeed, the linear behavior is only expected to be an asymptotic behavior at late times, at least for the non-interacting case.  The speed against the number of
particle extracted from a linear fit at times $t>16$ is shown in the inset of Fig.~\ref{fig:speed}. The chosen definition for the speed captures both the speed of the front $v^{}_{B}$
and its broadening, in such a way that we can not make definitive conclusion on the difference of speed $v^{}_{B}$ of the front.

In addition to the speed of front propagation, the behavior of the OTOC in the vicinity of the front is an interesting quantity that depends on the characteristics of the
system. In previous studies~\cite{Khemani_Huse_Nahum_2018,Xu_Swingle_2020}, the following universal form has been conjectured:
\begin{equation}
	\label{eq:fit_front}
	C^{}_{l\sim v^{}_{B}t}(t) \sim \exp\left(-c \frac{(x-v^{}_{B}t)^{1+p}}{t^{p}} \right),
\end{equation}
where $c$ is a constant, and $p$ is a system-specific exponent. It has been shown that $p=0$ for the Sachdev-Ye-Kitaev model~\cite{Gu_Qi_Stanford_2017}, $p=1/2$ for non-interacting
fermionic chains~\cite{Xu_Swingle_2020,Lin_Motrunich_2018a} and $p=1$ for interacting integrable
chains~\cite{Luitz_Lev_2017,Gopalakrishnan_Huse_Khemani_Vasseur_2018,Khemani_Huse_Nahum_2018} as well as for random circuit
models~\cite{Khemani_Vishwanath_Huse_2018,Rakovszky_Pollmann_Von_Keyserlingk_2018,Von_Keyserlingk_Rakovszky_Pollmann_Sondhi_2018,Nahum_Vijay_Haah_2018}. The random circuit models and
the interacting integrable chains hold the same coefficient as they both exhibit a diffusive front which belongs to the Kardar-Parisi-Zhang universality
class~\cite{Ljubotina_Znidaric_Prosen_2017,Nahum_Vijay_Haah_2018,Ljubotina_Znidaric_Prosen_2019}.  In Fig.~\ref{fig:fit_p} we show how the exponent $p$ varies with increasing particle
number. In Appendix~\ref{appendix:front}, we detail the fitting procedure that we used to extract the exponents $p$. For $n=1$ corresponding to the non-interacting fermionic case, we
recover the expected $p=1/2$. Then, it increases until it reaches the expected $p=1$ in the half-filled sector. These results hold for finite size systems: as the system size increases,
any non-extensive number of particle $n$ tends to behave as in the free case, while any particle number extensive with the system size ($n \simeq L/2$ or analogously any magnetisation 
$m$ close to $0$) is similar to the full interacting case. \\

\begin{figure}
	\includegraphics[width=1.\columnwidth]{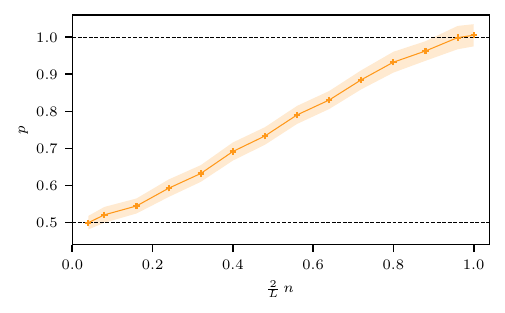}
	\caption{Exponent $p$ shown in Eq.~\eqref{eq:fit_front}, controlling the decay of the OTOC near the front against the number of particles $n$ for a chain of $L=50$
	sites. $p$ is fitted for the spatial decay after the front at different times, every other time between $t=4$ and $t=14$. The shaded region indicates the standard
	deviation of all the fits considered for each point, and the fitting protocol is detailed in Appendix~\ref{appendix:front}.}
	\label{fig:fit_p}
\end{figure}

\begin{figure}
		\includegraphics[width=1.\columnwidth]{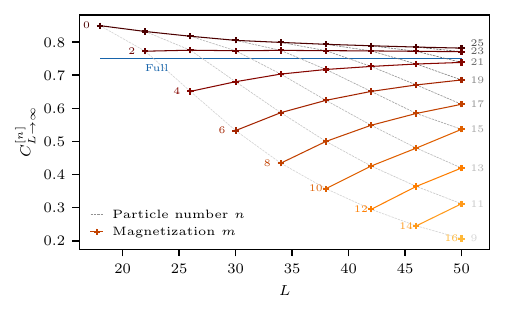} 
		\caption{$C_{\infty}$ in the thermodynamic limit in different sectors: $L\rightarrow \infty$ first, then $t\rightarrow \infty$. The points are obtained by fitting the
		data at intermediate times by a shifter power law, extracting the value at long times. Colored curves correspond to a fixed magnetization $m=L/2-n$, while grey dashed
		curves correspond to a fixed particle number $n$. The horizontal blue line is the expected value for the charge-preserving operator, which does not depend on system size.}
		\label{fig:TD_limit_value}
	\end{figure}

	\emph{Saturation:} After the front is reached, the OTOC oscillates and relaxes to a fixed value as $t\rightarrow \infty$. This regime can be observed in
	Fig.~\ref{fig:vertical_cuts} for $j=j'$ until the red circles, which indicate the beginning of the second OTOC growth due to boundary effects. As $L$ and $n$ increase, the
	oscillations are smoother and the OTOC appear to decay gradually toward a finite value greater than zero for $n>1$. Fig.~\ref{fig:TD_limit_value} illustrates saturation value
	$C^{[n]}_{L\rightarrow\infty}$ of the OTOC as $L$ increases for different projections $n$, obtained by fitting a shifted power-law decay. A blue horizontal line marks the
	saturation value of the full OTOC, which remains independent of system size. Both magnetization $m$ (colored solid lines) and particle number $n$ (dashed gray lines) are
	depicted.  In the thermodynamic limit, only the magnetization is relevant, as it is extensive when $m$ is finite around zero, whereas $n$ is more meaningful at fixed $L$ for
	highlighting deviations from the non-interacting case $n=1$. As the system approaches the thermodynamic limit, the curves corresponding to a fixed particle number converge to
	zero. Meanwhile, all fixed magnetizations approach the full symmetric operator value either from above, for $m\leq2$, or from below. This figure does not account for the
	factor $D_n/2^{L}$ from Eq.~\eqref{eq:otoc_decomposition}, focusing instead on the asymptotic behavior of each projected OTOC. In the thermodynamic limit, this factor
	becomes dominant in the half-filled sector, as $\lim\limits_{L\rightarrow \infty} D_n/2^{L} = \delta_{n,L/2}$, meaning that only the half-filled sector contributes to the full
	definition of the OTOC.  However, at finite $L$, all sectors have a non-zero contribution, leading the half-filled sector to overestimate the full OTOC.

	\subsubsection{Late times (finite size effects)}
\begin{figure}
		\includegraphics[width=1.\columnwidth]{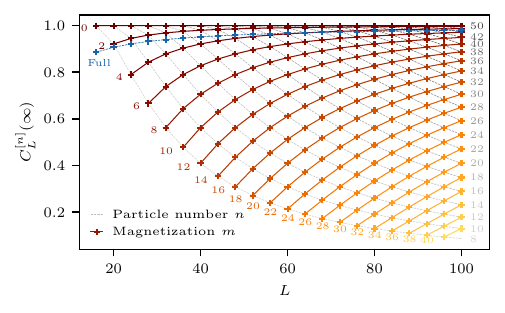}
		\caption{$C(\infty)$ in the finite size limit: $t\rightarrow \infty$ first, then $L\rightarrow \infty$. The points are obtained via the closed formula found through the
		Weingarten calculation. The legend is the same as in the previous plot, yet the expected value for the charge-preserving operator depends on system size.}
		\label{fig:Weingarten_scaling}
	\end{figure}

	When the information front reaches the open boundary, it gets reflected toward the other edge of the chain. This yields an extra scrambling of information resulting in second
	increase of the amplitude of the OTOC, as seen in Fig.~\ref{fig:vertical_cuts}. 
	After multiple reflections on the boundary, we expect that the operator loses all local information about its initial state and it is effectively rotated by a unitary random matrix:
	time-evolved local probes mimic Haar-random dynamics due to the non-conservation of $S_j^z$, even though the global dynamics remains non-Haar random.
	A similar treatment of the OTOC has been performed for their late time behavior in random circuits~\cite{Rakovszky_Pollmann_Von_Keyserlingk_2018}, where a local operator $\hat{V}$ 
	becomes $\tilde{V} = U^{\dagger} \hat{V} U$, with $U$ a Haar random unitary preserving a charge $\hat{Q}$. In our integrable system, we consider similarly random dynamics at late 
	times, and estimate the saturation value of the OTOC by taking the expectation value over the random unitary matrices $U$ by integration over the Haar measure.  In Appendix
	~\ref{appendix:weingarten}, we show the derivation of the late time saturation value of the OTOC using Weingarten calculus~\cite{Collins2010_Weingarten}, leading to the following result:
	\begin{equation}
		C^{[n]}_{L}(\infty) = 1 - \frac{2\mathrm{Tr}({\hat \sigma_j}^{[n]})^2}{D_n^{~2}} + \frac{\mathrm{Tr}({\hat \sigma_j}^{[n]})^{4}}{D_n^{~4}} .
		\label{eq:wg_otoc}
	\end{equation}
	This equation holds for the OTOC at $j'=j$, but can be obtained equivalently for each $j'$. The symmetry is protecting the full OTOC defined as in Eq.~\eqref{eq:otoc_decomposition} to vanish
	exponentially with system size as $2^{-2L}$, since the trace of ${\hat \sigma_j}^{[n]}$ is non zero everywhere outside the half-filled sector (for even system sizes). The expected saturation
	value is plotted in Fig.~\ref{fig:Weingarten_scaling} for increasing system sizes. In comparison to the thermodynamic limit saturation shown in Fig.~\ref{fig:TD_limit_value},
	the saturation value approaches $1$ algebraically for a finite deviation from the half-filled sector as $1-8{(L/2 - n)}^2L^{-2}$, which we obtain by expanding Eq.~\eqref{eq:wg_otoc} for $L\rightarrow \infty$.
	The half-filled sector $n=L/2$ is qualitatively different from the others, as it approaches $1$ exponentially fast with increasing system size as $1-L\exp(-2\log[2]L)\pi/2$ due to
	the cancellation of the trace in this sector. As also stated for the thermodynamic limit, any OTOC projected in a non-extensive particle number sector saturates to zero when
	$L\rightarrow \infty$.

\begin{figure}
		\includegraphics[width=1.\columnwidth]{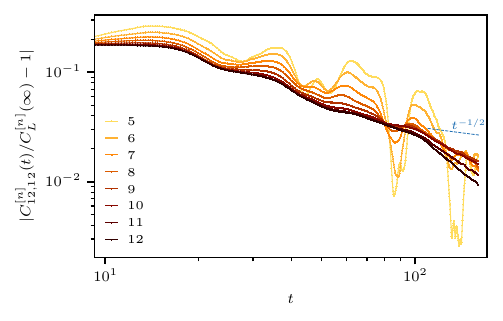}
		\caption{Convergence of OTOC in different sectors for $L=24$ towards the value $C(\infty)$ obtained by integrating the random unitary matrices. At late times, the
		convergence follows a power-law decay, above which a blue dashed line indicating a $t^{-1/2}$ power-law is plotted for a qualitative comparison.}
		\label{fig:convergence_Weingarten}
	\end{figure}

	The saturation values predicted using random unitary matrices are shown as dashed lines in Fig.~\ref{fig:vertical_cuts} for various projections and for the full OTOC. In
	Fig.~\eqref{fig:convergence_Weingarten}, we illustrate the convergence of the OTOCs toward the predicted value. To reduce the amplitude of oscillations, we plot the average value
	of the OTOC over a time window of $20$ time units, as function of time.  While the largest sectors converge smoothly to the value predicted in Eq.~\eqref{eq:wg_otoc}, the
	smaller ones exhibit stronger oscillations.  These oscillations are  due to the weaker effect of interactions in the small sectors, to which adds the method's approximation of
	the trace of the operator using the average of the expectation value over 5 Haar-random pure states~\cite{Luitz_Lev_2017,Colmenarez_Luiz_2020_Lieb_Robinson}.  As the dimension
	$D_n$ of the corresponding Hilbert space grows, the convergence is smooth and we are able to identify a power-law convergence to the predicted value at late times -- here
	$t>100$. We plot a blue dashed line showing the diffusive saturation of the OTOC, scaling as $t^{-1/2}$, in the thermodynamic limit of random circuit
	models~\cite{Rakovszky_Pollmann_Von_Keyserlingk_2018}. Here the OTOC in the late-time regime exhibits a faster convergence with an exponent bigger than $1/2$ in all the sectors
	shown: the number of crossing paths~\cite{Rakovszky_Pollmann_Von_Keyserlingk_2018} might substantially increase due to multiple boundary reflections, in contrast to the typical
	random walkers often used to model diffusive behaviors. \\

\section{Conclusion}
\label{Conclusion}

In this study, we extended the formalism of symmetric MPOs to allow the direct construction of MPOs projected onto a chosen irreducible representation of an abelian symmetry.  This was
achieved by generalizing the concept of operator charge through a parametrization of the supercharge operator with a single parameter, $\alpha$. Using this framework, we explored the
impact of interactions on the OTOC of an integrable Heisenberg spin chain through the number of particles $n$ in the system, connecting the non-interacting limit (sector with one
particle) to the fully interacting case at half-filling. 

We investigated the OTOCs in three different time regimes, from the very short times to times beyond the point where the front reaches the boundary, in which significant qualitative
behaviors and interaction effects are reported. At short times, interactions do not play a crucial role and the OTOCs in each sector exhibit the same power-law as predicted in the
non-interacting system.  After the first scrambling mechanism, the OTOC propagates in space and time within a light-cone, which spreads faster in sectors with more particle, due to a
larger widening of the front stemming from enhanced interactions. We showed that the free problem connects smoothly to the interacting one as the number of particle increases by
analyzing the decay of the OTOC near the front using previous analytical predictions~\cite{Khemani_Huse_Nahum_2018,Xu_Swingle_2020}. 

Finally, we examined the influence of open boundary conditions on the saturating value of the OTOC. While the operator entanglement entropy follows a volume law after reaching the boundary,
the MPO representation becomes ineffective. To address this, we employed an exact diagonalization method projected onto each sector to simulate the OTOC up to late-times for system
sizes up to $L\simeq 30$. As the initially local information carried by the OTOC reflects off the boundary multiple times, its value increases again before eventually saturating to a value
that we predict analytically. This prediction is made by replacing the time-evolution operator with a non-local Haar random unitary and integrating it out using Weingarten calculus.
This observation suggests that, while the system remains integrable, boundary conditions contribute to the full scrambling of local operators at long times. 
As a consequence, all local information about initial conditions are lost when interactions are introduced, whereas the non-interacting case continues to show strong oscillations. \\

\section{Acknowledgments}
\label{Acknowledgments}

The authors acknowledge support of the Deutsche Forschungsgemeinschaft through the cluster of excellence ML4Q (EXC 2004, project-id 390534769), the QuantERA II Programme that has received 
funding from the European Union’s Horizon 2020 research innovation programme (GA 101017733), the Deutsche Forschungsgemeinschaft through the project DQUANT (project-id 499347025), 
and the Deutsche Forschungsgemeinschaft through CRC1639 NuMeriQS (project-id 511713970) and CRC TR185 OSCAR (project-id 277625399). 

\bibliography{Biblio.bib}	

\clearpage
\begin{appendices}

\section{Scaling of the complexity of the projected MPO representation}
	\label{appendix:scaling}
    Here, we discuss an upper bound for the asymptotic scaling of the complexity of the different MPO representations considered, namely the non-symmetric, 
    symmetric full ($\alpha=-1$) and projected ($\alpha=L+1$) MPOs. We investigate the complexity through the scaling of the cost of the SVDs at the middle of the 
    chain, which is defined as
	$C_{\mathrm{SVD}}=M^{2}\times N$, where $M$ and $N$ are the dimensions of the matrix being decomposed, with $M<N$. In the middle of the chain, we always have $N=M$.  

	\emph{Non-symmetric MPO}: the bond dimension of the MPO grows by a factor $2^{2}$ at every site, as the tensors that we consider represent single half-spin particles. In the
	middle of the chain, there is a single matrix such that:
	\begin{equation}
		N = 2^{L}, \quad C=2^{3L}.
	\end{equation}

	\emph{Symmetric full MPO}: the bipartition, into subsystem A and subsystem B, in the middle of the chain is now composed of multiple \textit{internal sectors} with charges 
	$q^{}_{A}=n_{A}-n'_{A}\in[-L/2,L/2]$, on which SVDs can be independently performed (we use $q\equiv q^{}_{A}$ to simplify the notation). 
	The global cost is thus the sum of costs of each internal sector, which are defined from their dimensions as: \begin{equation}
		\begin{split}
			N^{}_{q} &= \sum\limits_{t=0}^{L/2 - |q|}\frac{\frac{L}{2}!}{t!\left( \frac{L}{2} - t\right)!}\frac{\frac{L}{2}!}{\left(|q|+t\right)!  \left(
			\frac{L}{2}-|q|-t\right)!},\\
			C &= \sum\limits_{q=-L/2}^{L/2} N_{q}^{3}.
		\end{split}
	\end{equation}
	The calculation of $N_{q}$ can be understood as the sum (all sub-vector spaces holding a charge $q$) of the product (because of the vectorization of the bipartite operator) of
    both dimensions of the blocks lying on the $q$-th diagonal (upper or lower 
    for $q$ respectively positive or negative) of the matrix representation of an 
    operator for which the
	subsystem B (second half of the chain) has been traced out.  

	\emph{Projected MPO}: the internal charge is now defined by $q=n_{A}+(L+1)n'_{A}$, such that each doublet $(n_{A}^{},n'_{A})$ represents a single symmetry sector. The
	vector space with $n_{A}$ particles in the half-chain has the usual dimension $N_{n^{}_{A}}^{}$ defined in the equation below, such that the dimension of the sector $q$ is
	$N_q = N_{n^{}_{A}}^{} N_{n'_{A}}^{}$.  Due to the boundary conditions defining the projection, only a part of the possible internal sectors is spanned. The allowed internal
	sectors $q$ for a given projection in the sector $n$ lies in the interval $[n_{\min},n_{\max}] = [\max(0,n-L/2), \min(L/2, n)]$. The cost for the sector representation is
	thus defined as:
	\begin{equation}
		\begin{split}
			N^{}_{n^{}_{A}} &= \frac{\frac{L}{2}!}{n^{}_{A}!\left(\frac{L}{2}-n^{}_{A}\right)!}, \\
			C^{[n]}  &= \sum\limits_{n^{}_{A},n'_{A}=n_{\min}}^{n_{\max}}N^{}_{n^{}_{A}} N^{}_{n'_{A}} = \left(\sum\limits_{n^{}_{A}=n_{\min}}^{n_{\max}}
			N_{n^{}_{A}}^{3}\right)^{2}.  
		\end{split}
	\end{equation}
	The cost of computing the full operator is obtained by summing over the cost of every projection, $C=\sum_{n=0}^{L} C^{[n]}$. \\

\begin{figure}
		\includegraphics[width=1.\columnwidth]{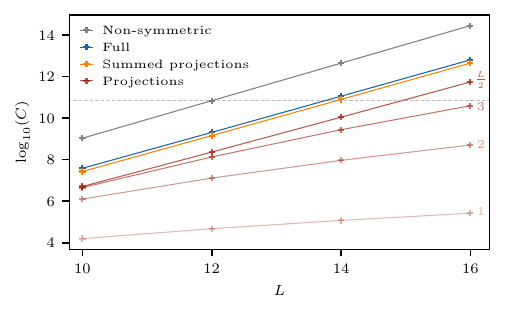}
		\caption{Predicted scaling of the computational cost for non-symmetric, symmetric full and projected MPO representations. The summed cost for the projected
		representation is plotted for comparison with the symmetric full MPO, corresponding to the same information about the operator. Only the projections
		$n=1,2,3,\frac{L}{2}$ are plotted for visibility. The dashed horizontal line corresponds to the cost of a non-symmetric calculation with $L=12$.} 
		\label{fig:predicted_scaling}
	\end{figure}

	In the discussion above, note that no physical arguments were used, and only the worst case is considered. In practice, operators have structure which lowers the
	predicted cost. The scaling of the different representations is shown in Fig.~\ref{fig:predicted_scaling}.

\section{Benchmarking projected MPOs}
	\label{appendix:benchmark}
	\begin{figure}
		\includegraphics[width=1.\columnwidth]{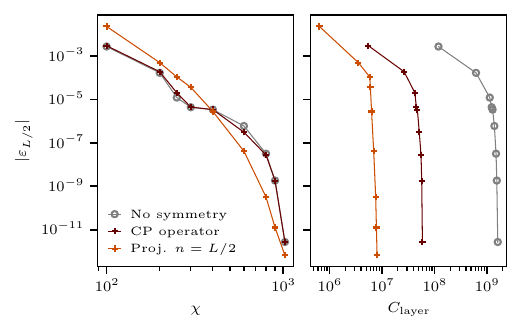}
		\caption{Truncation error for non-symmetric, charge-preserving and projected representations of the time-evolution operator $U(t=5)$ for $L=10$ and $dt=1/6$. Using a $4^{\mathrm{th}}$
		order Trotter decompositions, the corresponding errors are of the order of the machine precision. The MPOs are represented as matrices and are projected in the
		half-filled sector in order to compute $\varepsilon$. \textit{Left}: Error against $\chi$, defined as the number of kept states among all the sectors.  \textit{Right}:
		Error against total cost $C_{\mathrm{layer}}$, corresponding to the sum of the cost $C$ of each SVD during the last layer of Trotter gates.}
		\label{fig:benchmark_chi}
	\end{figure} 
	
	In the first appendix, we discussed the complexity of the different MPO representations in the situation where each block allowed by the symmetry is filled. We turn now to an
	actual situation by considering the simulation of the time-evolution operator of a 1d spin chain. Beyond the computational complexity, it is especially important to control the
	convergence of our computations. In general, a great deal of importance is attached to the bond dimension, being directly related to the convergence of the algorithm, at least
	in one-dimensional systems where a canonical form can be drawn. The bond dimension $\chi$ refers to the number of kept singular values of the operator's reduced density matrix
	$\rho^{}_{A}$ given a bipartition $(A,B)$. For symmetric representations, the reduced density matrix is built by sectors, and $\chi$ is defined as the sum of kept singular
	values in each sector such that $\chi=\sum_{q^{}_A} \chi_{q^{}_A}$. 

	As in the main text, we consider the isotropic Heisenberg model with open boundary conditions:
	\begin{equation}
		H = \sum_{i=1}^{L} \frac{J}{2}\left( \sigma^{+}_{i} \sigma^{-}_{i+1} + \sigma^{-}_{i} \sigma^{+}_{i+1}\right) + J \, \sigma^{z}_{i}\sigma^{z}_{i+1},
		\label{eq:app_Heisenberg}
	\end{equation}
	which we time evolve until a time $t$ using a Trotterized version of the time evolution operator $U(t)=\exp(-i H t)$ using the TEBD algorithm adapted for MPOs.  Unless otherwise
	stated, we choose $J=1$ and $t=5$, for a system of $L=10$ sites for which the contraction of an MPO into a matrix is tractable. Every dimensional quantity are implicitly taken
	in units of $J$. We checked that using a $4$-th order Trotter evolution with a time step of $dt=t/30$ allows to converge the time-evolution operator $U(t)$ up to machine
	precision (in terms of Froebenius norm). 

	We now compare the ability of the different representations to correctly reproduce the operator $U$ projected in the half-filled sector, which is relevant for MPO-MPS simulation
	with the MPS being in this sector (e.g. DMRG ground-state calculations). We consider the following error for comparison:
	\begin{equation}
		\eps^{\pd}_{n} = 1-\frac{\mathrm{Tr}\left[\left(U^{[n]}_{\mathrm{exact}}\right)^{\dagger} U^{[n]}_{\mathrm{MPO}}\right]}{\mathcal{D}_{n}^{\pd}},
		\label{eq:eps_sect}
	\end{equation}
    where $\mathcal{D}_{n}=\mathrm{Tr}\left(\mathds{1}^{[n]}\right)$ is the dimension of 
    the sector of charge $n$, and $U^{[n]}$ is the evolution operator at time 
    $t$ projected in this same sector. The
	absolute value of the error is plotted against the bond dimension defined as above in the left panel of Fig.~\ref{fig:benchmark_chi}, for the three MPO representations. The
	absolute value is taken to avoid the spurious negative values of $\eps$ for some $\chi$ in the $\alpha=-1$ symmetric and non-symmetric representations. These representations
	are only constrained to have their trace to sum up to a value below $2^{L}$, but the trace in each sector can exceed their dimension because of truncation.

	We observe in Fig.~\ref{fig:benchmark_chi} a steady decrease of the error for the projected representation, unlike the others. Each additional kept state in the projected MPO
	contributes to the sector of interest and hence participates to reducing the error. On the contrary, states added in other representations might contribute to other sectors
	also, in such a way that $\varepsilon_n$ does not decrease smoothly.

	Usually, the bond dimension is referred to as a indicator of the computational effort, improving the accuracy as we increase the effort. The internal structure of tensors
	changes the way of relating bond dimension and computational effort, since several distributions of states in sectors lead to different run-times and accuracies. To alleviate
	this issue, we examine another measure of the computational effort related to the cost of SVDs over a single Trotter step $C_{\mathrm{layer}}$ (chosen to be the last one), which
	consists for the Hamiltonian in Eq.~\eqref{eq:app_Heisenberg} in two layers of two-body gates starting from odd then even sites. Within these two layers, each SVD adds a cost
	$N\times M^{2}$ (where $N>M$, for an $(N\times M)$ dimensional matrix) to the total cost $C^{}_{\mathrm{layer}}$. The error in Eq.~\eqref{eq:eps_sect} is plotted against
	$C^{}_{\mathrm{layer}}$  in the right panel of Fig.~\ref{fig:benchmark_chi}. For the most accurate computations (and thus most expensive), we retrieve the expected scaling of
	the cost presented in Fig.~\ref{fig:predicted_scaling} for $L=10$. 

	\begin{figure}
		\includegraphics[width=1.\columnwidth]{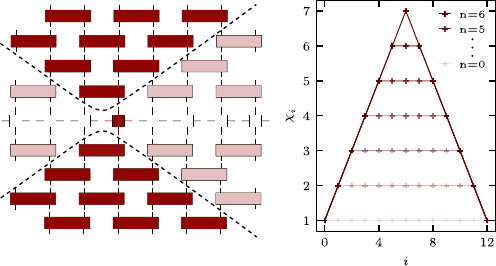}
		\caption{\emph{Left}: Sketch of the time-evolution of a local operator through a Trotter circuit. The expected information light-cone is represented by the
		dashed lines in both time directions, where entanglement builds up. Outside this part of the circuit, the operator remains trivial and transparent gates cancel each
		other out. \emph{Right}: Bond dimension along the chain at time $t=0$ of any Pauli string projected in sector $n$. The corresponding bond-dimension encodes
		classical information, related to the different internal sectors, while no quantum entanglement is present.}
		\label{fig:internal_structure}
	\end{figure} 
	\textit{Time evolution of local operators:} In order to investigate OTOCs or the entanglement structure of local operators using tensor networks, the operator are defined in the Heisenberg
	picture as $\hat{O}(t) = U^{\dagger}(t) \hat{O} U(t)$. Generally, the MPO at the initial time has a bond dimension $\chi=1$ along the whole chain. As shown
	in the left panel of Fig.~\ref{fig:internal_structure}, the TEBD algorithm is easily to operators adapted by performing the sequence of Trotter gates on the ingoing legs as well
	as on the outgoing ones. For local operators, information (closely related to entanglement) spread within the light-cone in both directions, such that gates outside of this
	region simply cancel each other $U^{\dagger}_{n}U^{}_{n}$, as we consider a unitary time evolution. Hence, most of the tensors forming the MPO remain trivial with a bond
	dimension $\chi=1$ until the light-cone reaches them, yielding lighter simulations for short times. 

	In the case of projected MPOs, constraints affect the internal structure giving rise to an increase of the bond dimension along the chain. The bond dimension depends on the
	chosen projection, being maximal for the half-filled sector as shown in the right panel of Fig.~\ref{fig:internal_structure}. However, this increase does not affect the
	cancellation of gates outside the light-cone, as the internal structure only encodes classical information, and the quantum entanglement is still confined within the light-cone.

\section{Exact derivation of OTOC for the \texorpdfstring{$XX$}{XX} spin chain}
	\label{appendix:free_model}
	Finding a closed form for the one-particle sector of the $XX$ spin-$1/2$ chain is possible through a mapping to non-interacting fermionic degrees of freedom, since the
	density-density-like terms $S^{z}_{i}S^{z}_{i+1}$ in the Hamiltonian only brings an energy shift in this situation. The corresponding spin model is the $XX$ spin chain, defined
	as: 
	\begin{equation}
		\label{FreeHeis}
		\hat{H}_{XX} =  \frac{J}{2} \sum_i (\hat{\sigma}_i^{~+} \hat{\sigma}_{i+1}^{~-} + \hat{\sigma}_i^{~-} \hat{\sigma}_{i+1}^{~+} ).
	\end{equation}
	It can be mapped to fermionic degrees of freedom through the Jordan-Wigner (JW) transformation: $\hat{\sigma}_{i}^{+}=\prod_{j=1}^{i-1}e^{i c^{\dagger}_{j}c^{}_{j}}
	\hat{c}_i^{\dagger}$, which transforms Eq.~\eqref{FreeHeis} into:
	\begin{equation}
		\label{HfreeFerm}
		\hat{H}=\frac{J}{2} \sum_j^{L}\left(\hat{c}_{j}^{\dagger} \hat{c}^{}_{j+1}+\hat{c}_j^{\dagger} \hat{c}^{\pd}_{j+1}\right),
	\end{equation}
	with $\hat{c}^{\dagger}_{i}$ (resp. $\hat{c}_{i}$) creating (resp. annihilating) a fermion at site $i$. In order to derive a simple closed analytical form, we use periodic boundary
	conditions for the chain, which is then diagonalized by the rotation in Fourier space: $\hat{c}^{}_j=1/\sqrt{L} \sum_k e^{i k j}
	\hat{c}^{}_k$, where $k=2\pi n/L$, $n\in\{0,\cdots,L-1\}$, are the pseudo momenta. The spectrum of the Hamiltonian~\eqref{HfreeFerm} is the usual $\eps_k^{}=J\cos(k)$. In order
	to write the OTOC in the Heisenberg representation, we write the time evolution operator in the eigenbasis:
	\begin{equation}
		\hat{U}(t)=e^{-i \hat{H} t} = \prod_{k} e^{-i J t \sum_{k} \cos(k) \hat{c}_{k}^{\dagger} \hat{c}^{\pd}_{k}}.
		\label{eq:time_evol_op}
	\end{equation}
	Inserting this form of $U$ and the JW for $\hat{\sigma}^{z}_{j}$: $\hat{\sigma}_{i}^{z}  =2 \hat{n}_i - \mathds{1}$ into the definition of OTOC~\eqref{eq:def_otoc}, we obtain:
	\begin{equation}
		\label{simpl_freeOTOC}
		\mathcal{C}_{j, j^{\prime}}\left(t\right) = 2^{5} \left( \mathrm{Tr}\hspace{-0.06cm}\left[\hat{n}^{}_{j}(t) \hat{n}^{}_{j'} \hat{n}^{}_{j}(t) \hat{n}^{}_{j'}\right] -
		\mathrm{Tr}\hspace{-0.06cm}\left[\hat{n}^{}_{j}(t) \hat{n}^{}_{j'}\right] \right).
	\end{equation}
	Using the fact that the system only contains one particle, we can use the following commutation relation to simplify equations:
	\begin{equation}
		\hat{c}^{\pd}_{k} \hat{c}_{k'}^{\dagger} \hat{c}^{\pd}_{k''}=\left(\delta_{k k'}^{\pd}-\hat{c}_{k'}^{\dagger} \hat{c}^{\pd}_{k}\right) \hat{c}_{k''}^{\pd} = \delta_{k
		k'}^{\pd}\hat{c}_{k''}^{\pd}.
	\end{equation}
	After developing Eq.~\eqref{simpl_freeOTOC} with the commutation relations shown above, one arrives at the following result:
	\begin{equation}
		\label{exactOTOC_discrete}
		\begin{split}
			\mathcal{C}_{j, j^{\prime}} = &\frac{32}{L^{2}} \sum_{k k^{\prime}} e^{i\left(k-k^{\prime}\right) j} e^{i J t(\cos (k) -\cos (k^{\prime}))} \\
			- &\frac{32}{L^{4}} \left(\sum_{k k^{\prime}} e^{i\left(k-k^{\prime}\right) j} e^{i J t(\cos (k) -\cos (k'))}\right)^{2}
		\end{split}
	\end{equation}
	In order to evaluate the sums over $k,k'$, we consider the thermodynamic limit $L\rightarrow \infty$ as follows:
	\begin{equation}
		\frac{1}{L}\sum\limits^{L-1}_{n=0} \rightarrow \int\limits^{1}_{0}\mathrm{d}x, \quad \frac{n}{L} \rightarrow x,
	\end{equation}
	By introducing further the following change of variable: $\kappa = (x-y)/2, \quad \theta=(x+y)/2-\pi$, the integral to solve reads:
	\begin{equation}
		\begin{split}
			I &= \frac{1}{2\pi^{2}}\int\limits^{\pi}_{-\pi}\mathrm{d}\kappa \,e^{i 2\kappa (j-j')}\int\limits^{\pi}_{-\pi}\mathrm{d}\theta \, e^{i J t \sin (\theta) \sin(
			\kappa )} \\
			&= \frac{1}{2\pi}\int\limits^{\pi}_{-\pi}\mathrm{d}\kappa \, e^{i 2\kappa (j-j')} J^{}_{0}(i J t \sin( \kappa )),
		\end{split}
	\end{equation}
	where $J_{0}(x)$ is the Bessel function of first kind of integer order $0$. We expand the Bessel function in series, and recall that for OTOC, we only need the real part
	(cosine part of the exponential). We thus obtain the following result:
	\begin{equation}
		\begin{split}
			I &= \frac{1}{2\pi}\sum\limits^{\infty}_{k=0} \frac{(-1)^{k}(2Jt)^{2k}}{4^{k}(k!)^{2}} \int\limits^{\pi}_{-\pi}\mathrm{d}\kappa \, \cos (i 2 l\kappa) \sin(\kappa)^{2k}\\
			&= (-1)^{l}\sum\limits^{\infty}_{k=0} \frac{(-1)^{k} \Gamma(1+2k)(\frac{1}{4}(Jt)^{2})^{k}}{(k!)^{2} \Gamma(1+k-l) \Gamma(1+k+l)},
		\end{split}
	\end{equation}
	where $l=j-j'$ denotes the spatial extent from the initial point $j$. We recognize the definition of $J_{\nu}(z)J_{\mu}(z)$ given in Eq.~(9.1.14) of the textbook of Abramowitz and
	Stegun~\cite{abramowitz_stegun}, identifying $\nu=l$ and $\mu=-l$ and using $k! = \Gamma(1+k)$. 
	We thus obtain: 
	\begin{equation}
		I = (-1)^{l} J_{-l}(Jt) J_{l}(Jt) = J_{l}^{2}(Jt),
	\end{equation}
	using the property of integer order Bessel functions that $J_{-l}(x)=(-1)^{l}J_{l}(x)$.  We finally arrive at the very simple result \eqref{eq:exact_OTOC} presented in the text
	for the OTOC of free fermions in the large $L$ limit for periodic boundary conditions.

\section{Early time growth of OTOC}
	\label{appendix:early_times}
	\begin{figure}
		\includegraphics[width=1.\columnwidth]{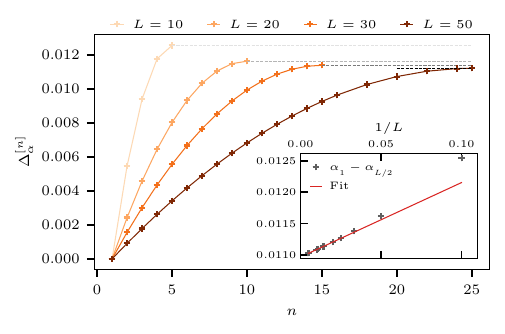} 
		\caption{Comparison of the power-law exponent at early times to the one-particle sector $\Delta_{\alpha}^{[n]}=|\alpha^{[n]}_{1}-\alpha^{[1]}_{1}|$, for different system
		sizes. The spatial separation of operators is $l=1$. \emph{Inset}: Finite size scaling of the largest difference $|\alpha^{[L/2]}-\alpha^{[1]}|$. The fitted data follows
		the following equation: $\Delta_{\alpha} = 0.011 + 0.015\times L^{-1}$, indicating that the difference remains (although it is small) in the thermodynamic limit. For such
		early times $t<1$, we reach systems of up to $L=200$ sites in the half-filled sector to get the finite size scaling.} 
		\label{fig:early_times_scaling}
	\end{figure}

	At early times, the OTOCs are expected to grow as a power law, as one can observe in Fig.~\ref{fig:early_times_otocs}. We use the closed form Eq.~\eqref{eq:exact_OTOC} for OTOC
	in the one-particle sector to derive the corresponding power law $t^{\alpha_l^{}}$, which depends on the spatial separation $l=|j'-j|$. To extract the exponent, we
	calculate the following quantity: 
	\begin{equation}
		\label{powerlaw}
		\frac{\partial}{\partial t} \left( \log (C^{}_{i,l}) \right) \propto \frac{\partial}{\partial t} \left( \alpha^{}_{l}\log (t) \right) = \frac{\alpha^{}_{l}}{t}.
	\end{equation}
	Taking the derivative of the log, one has to take the derivative of Eq.~\eqref{eq:exact_OTOC} with respect to time. We use the following definition for the derivative of Bessel
	functions of order $l$:
	\begin{equation}
		\partial_{t} J_{l}(t) = J_{l-1}(t) - \frac{l}{t}J_{l}(t).
		\label{der_bess}
	\end{equation}
	We insert Eq.~\eqref{der_bess} into Eq.~\eqref{powerlaw}, and find the closed form for the derivative of OTOC at time $t$ for the free case valid at all times $t$:
	\begin{equation}
		\frac{\partial \log(C^{}_{i,l}(t) )}{\partial t} = 2 \frac{\left(J^{}_{l-1}(t) - \frac{l}{t}J_{l}(t)\right)\left(1-2J^{2}_{l}(t)\right)}{J^{}_{l}(t)(1-J^{2}_{l}(t))}.
	\end{equation}
	We are interested in the limit of early times to get the exponent $\alpha_{l}$ when $t\rightarrow 0$. The Taylor expansion of Bessel functions reads: 
	\begin{equation}
		\label{taylor_bessel}
		J_{l}(t) = \frac{t^{l}}{2^{l}\Gamma(l+1)}\left(1 - \frac{t^{2}}{4(l+1)} + ... \right).
	\end{equation}
	By considering the expansion up to $t^2$, we obtain the following form for the exponent $\alpha^{}_{l}$:
	\begin{equation}
		\alpha_l \approx 2l\left(1-\frac{1}{l\left(\frac{4(l-1)}{t^{2}}-1\right)}\right).
	\end{equation} 
	It is clear that in the limit $t\rightarrow 0$, the first order gives $\alpha_{l} = 2l$, as found by perturbation of the Baker-Campbell-Hausdorff formula~\cite{Colmenarez_Luiz_2020_Lieb_Robinson,Riddell_Sorensen_2019_rand_XX,Fortes_GarciaMata_Jalabert_Wisniacki_2019}.

	In the main text, it was shown that the exponent of the sectors with $n>1$ deviates from the value of the one-particle sector. The deviation being small in front of the absolute
	value of the exponent, we show that they remain finite in the thermodynamic limit in Fig.~\ref{fig:early_times_scaling}. However, this discussion concerns distinctions of the
	power-laws at finite time $\delta t$ due to the numerical discretization of time, while the result
	of~\cite{Colmenarez_Luiz_2020_Lieb_Robinson,Riddell_Sorensen_2019_rand_XX,Fortes_GarciaMata_Jalabert_Wisniacki_2019} indicates that at $t=0$ each sector $n$ carries the same
	exponent.

\section{Front propagation}
	\label{appendix:front}
	\begin{figure}
		\includegraphics[width=1.\columnwidth]{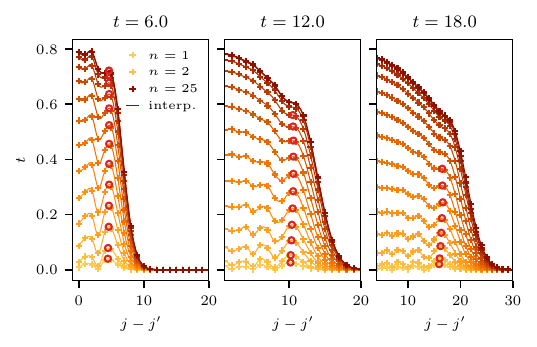}
		\caption{Spatial region around the front of OTOC at different times, for all sectors $n$ with $L=50$ sites. The initial local operator is located at site $j=15$. The
		solid lines are obtained through an interpolation of the actual data, marked by pluses. Reds circles indicates the first maxima from the end of the chain, indicating the
		end of the information front. After a certain time, the largest sectors do not show any maximum anymore.}
		\label{fig:interpolated_otocs}
	\end{figure}

	In this appendix, we discuss the propagation of the information front in the OTOC in different sectors as plotted in Fig.~\ref{fig:speed}. The front corresponds to the region
	in space where the OTOC starts decaying on an exponential scale at a given time. In order to extract the front at each time step, we interpolate the OTOC in space to observe a
	smooth variation of the front, as one can see in Fig.~\ref{fig:interpolated_otocs}. Ideally, the front would be defined as the first maximum of the OTOC from the end of the
	chain, which is indicated for different times by red circles in Fig.~\ref{fig:interpolated_otocs}. However, the front is not always well-defined, as the OTOC in the largest
	sectors does not show any maximum anymore after a certain time, such that we have to identify the front differently. As the decay is always well-defined before reaching the
	boundaries, we choose to define the front as the point where the OTOC reach 1\% of their maximal value at this given time $t$, which indicates that the information started
	decaying exponentially and that we are leaving the light-cone. 

	\begin{figure}
	    \includegraphics[width=1.\columnwidth]{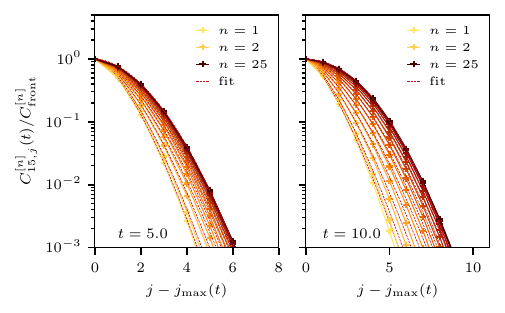}
	    \caption{Decay of the OTOC after reaching the front for different projections, for the same parameters as the previous plots. Each curve is normalized to its value at the
	    front $C_{i,j_{\mathrm{max}}}(t)$ for visibility. \emph{left}: $t=5$. \emph{right}: $t=10$.}
	    \label{fig:decay_after_front}
	\end{figure}

	We discuss now the behavior of the OTOC near the front. In Fig.~\ref{fig:fit_p} of the main text, we plot the fitted exponent $p$ of Eq.~\eqref{eq:fit_front} as function of the
	number of particles $n$ in the sector. The exponent $p$ is independent of the time at which the front is reached, and is only dependent on the sector $n$. However, as the front
	broadens in time, it is harder to extract its position at later times, and the discrete time and space grids makes it even more difficult. Hence, we fit each sector at $10$
	different times between $t=4$ and $t=14$ and use the average value and its standard deviation as the final result shown in Fig.~\ref{fig:fit_p}. In
	Fig.~\ref{fig:decay_after_front}, we show the decay of the OTOC around the front for two different times and the corresponding fits. The OTOC in each sector is normalized to its
	value at the front for visibility, such that we can compare their spatial propagation. Note that the times at which the front is reached differs slightly in each sector, and the
	time presented in the figure serves as an indicator.

\begin{figure}
		\includegraphics[width=1.\columnwidth]{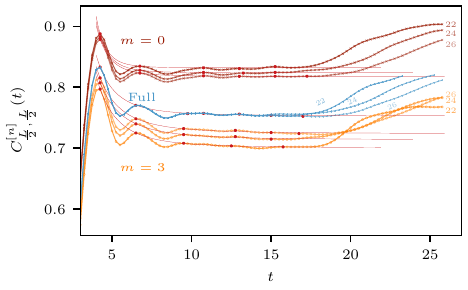}
		\caption{The full OTOCs for different system size exhibit the same behavior and differ for the time at which they start increasing.  For different $L$, are shown OTOCs
		with the same magnetization $m = L/2 - n$, here $m=0$ and $m=3$.}
		\label{fig:otoc_fits}
	\end{figure}

	The last point that we tackle here is the dependence of the OTOC saturation value in the thermodynamic limit with system size. In Fig.~\ref{fig:otoc_fits}, we show the OTOCs
	projected in different magnetization sectors for different system sizes, as well as the full OTOC (sum of all sectors). Before the boundaries are reached, we don't expect any
	dependence of the full OTOC on the system size, as the local operator can spread in space freely. However, it is not the case for the sector projected ones: as the system size
	increases, more sector are available, and each sector has to share the information with the others to keep the full OTOC unchanged. These two effects are clearly observed in
	Fig.~\ref{fig:otoc_fits}. In the thermodynamic limit $L\rightarrow \infty$, only the contribution of the largest sector with magnetization $0$ remains, which should then be
	equivalent to the full OTOC. As seen in both Fig.~\ref{fig:otoc_fits} and Fig.~\ref{fig:TD_limit_value}, the $m=0$ sector converges to the full saturation value as the system
	size increases. The other sectors (of fixed magnetization, any fixed particle number $n$ will see its saturation value vanish in the thermodynamic limit) also appear to converge
	to the full OTOC, but it is not relevant as they contribute with a factor $D_n/2^{L}$ to the full OTOC, which vanishes in the thermodynamic limit for $n\neq L/2$.
	
\section{Late times: OTOC saturation value for a random unitary time-evolution}
	\label{appendix:weingarten}
	In this appendix, we detail the derivation of the saturation value of OTOC $C_{L}(\infty)$ as shown in Eq.~\eqref{eq:wg_otoc}. We consider that the time-evolution of local
	operators $S^{z}(t\rightarrow \infty)$ is similar to a random rotation, such that information is lost about initially local properties of the operator. Starting with the
	definition Eq.~\eqref{eq:def_otoc}, we consider the averaged value over the Haar measure of the $d$-dimensional unitary group $U(d)$:
	\begin{equation}
		\mathbb{E}_{U}(Z^{[n]}) = \int \mathrm{d}U \, \mathrm{Tr}\left[ U^{\dagger}Z^{[n]} U Z^{[n]} U^{\dagger}Z^{[n]} U Z^{[n]} \right],
		\label{eq:weingarten_expect}
	\end{equation}
	where $Z^{[n]}=(\sigma^{z})^{[n]}$ is $Z$-Pauli operator projected in the sector of $n$ particles, and $C_{L}(\infty) = 1-\mathbb{E}_{U}(Z^{[n]})/D_n$. The average over the
	unitary group is performed using Weingarten calculus, using its graphical representation. Following~\cite{Collins2010_Weingarten}, we can express the expected value in terms of
	diagrams $\mathcal{D}_r$ obtained after a removal procedure $r$ of the original diagram $\mathcal{D}$, chosen in the space of possible removals $r\in \mathrm{Rem}(\mathcal{D})$.
	The expectation value thus reads:
	\begin{equation}
		\mathbb{E}_{U}(\mathcal{D}) = \sum_{r=(\alpha,\beta)\in\mathrm{Rem}(\mathcal{D})}\mathcal{D}_r \mathrm{Wg}(d, \alpha\beta^{-1}), \
		\label{eq:removals}
	\end{equation}
	where  $\mathrm{Wg}(d, \sigma)$ is the Weingarten function, $d$ is the dimension of the matrix $U$, and $(\alpha,\beta)$ is a set of two permutations defining the corresponding
	removal $r$, which we explain further below. In the large-$d$ limit, the Weingarten function takes the following form:
	\begin{equation}
		\begin{split}
			\mathrm{Wg}(d, \sigma) &= (-1)^{p-\#\sigma}\hspace{-0.4cm}\prod_{C\in \mathrm{Cycles}(\sigma)}\hspace{-0.4cm}\mathrm{Wg}(d,C)(1+\mathcal{O}(n^{-2})) \\
			\mathrm{Wg}(d, C) &= (-1)^{|C|-1}c_{|C|-1}^{} \hspace{-0.4cm}\prod_{-|C|+1\leq j \leq|C|-1}\hspace{-0.4cm} (d-j)^{-1}.
		\end{split}
	\end{equation}
	$\# \sigma$ is the number of cycles in the permutation $\sigma$, $p$ is the number of elements in the permutation, $|C|$ the length of the cycle $C$ and $c_{i}$ is the $i$-th
	Catalan number $c_{i} = (2i)! / ((i+1)!i!)$ \\
	\begin{figure}
		\includegraphics[width=0.6\columnwidth]{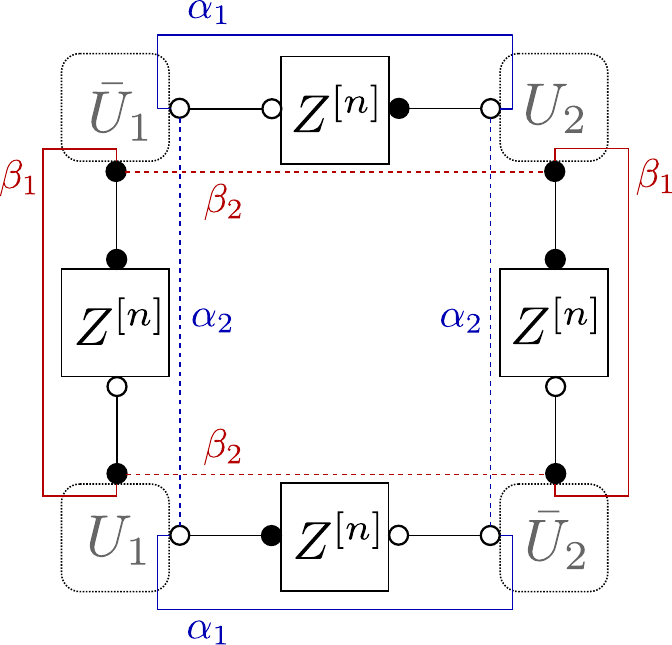}
		\caption{Weingarten diagram for the trace of operators $\mathrm{Tr}\left(U^{\dagger}S^{z}_{i}U S^{z}_{i}U^{\dagger}S^{z}_{i}U S^{z}_{i}\right)$. We note $U^{\dagger} =
		\bar{U}^{t}$, where the transpose is performed by inverting the direct and dual spaces, denoted by empty and full circles respectively. Blue lines connect direct spaces
		of $U_{i}$ and $U_{\alpha(i)}$ and red lines connect the dual spaces of $U_{i}$ and $U_{\beta(i)}$. Solid and dashed lines correspond to the two different possible
		permutations for $\alpha$ and $\beta$.}
		\label{fig:weingarten_diagram}
	\end{figure}
	\emph{Removals}: In order to explain the way to perform the removal, we refer to Fig.~\ref{fig:weingarten_diagram}, which represents the current calculation of interest
	Eq.~\eqref{eq:weingarten_expect}. In this diagram, white (resp. black) circles refer to the direct (dual) Hilbert space on which an operator (represented by boxes) act. We also
	performed the transpose of $\bar{U} = (U^{\dagger})^{t}$ by inverting black and white circles on the corresponding operators. In this case, $U_{1,2}$ and $\bar{U}_{1,2}$
	correspond to a same random realization of a random unitary matrix $U$, but the indices are useful to define a removal $r=(\alpha, \beta)$. $\alpha$ (resp. $\beta$) is a
	permutation connecting the white (black) circle of $U_{i}$ to the white (black) circle of $\bar{U}_{\alpha(i)}$ ($\bar{U}_{\beta(i)}$). Each combination of $\alpha$ and
	$\beta$ corresponds to a removal $r$: we connect the lines (vector spaces) according to the chosen permutations, and read what is left in $\mathcal{D}_r$. In
	Fig.~\ref{fig:weingarten_diagram}, we have two choices for $\alpha$ (blue lines) and two for $\beta$ (red lines), which leads to four reduced diagrams. We summarize the
	different combinations in the following table:
	\begin{equation*}
		\begin{matrix}
			\alpha 	&&	\beta  &&  \mathcal{D}_r 				&& \mathrm{Wg}(d, \alpha\beta^{-1})\\
			\hline
			(12) 	&&	(1)(2) &&  \mathrm{Tr}(Z)^{4}				&& -(d(d-1)(d+1))^{-1} \\
			(12) 	&&	(12)   &&  \mathrm{Tr}(Z^{2})\mathrm{Tr}(Z)^{2}		&& (d^{2})^{-1} \\
			(1)(2) 	&&	(1)(2) &&  \mathrm{Tr}(Z^{2})\mathrm{Tr}(Z)^{2}		&& (d^{2})^{-1} \\
			(1)(2) 	&&	(12)   &&  \mathrm{Tr}(Z^{2})^{2} 			&& -(d(d-1)(d+1))^{-1}
		\end{matrix},
	\end{equation*}
	where $Z$ implicitly refers to $Z^{[n]}$. While the trace of $Z^{[n]}$ is non-zero in sectors $n\neq L/2$, it is always true that $(Z^{[n]})^{2}=\mathds{1}$, such
	that $\mathrm{Tr}(Z^{2}) = D_{n} = d$. We can thus derive the final form through Eq.~\eqref{eq:removals}:
	\begin{equation}
		C^{[n]}_{L}(\infty) = 1 - \left(\frac{2\mathrm{Tr}(Z^{[n]})^2}{d^{2}}-\frac{d^{2} + \mathrm{Tr}(Z^{[n]})^{4}}{d^{2}(d-1)(d+1)}\right),
		\label{eq:wg_res}
	\end{equation}
	with:
	\begin{equation}
		d = \frac{L!}{(L-n)!n!}, \quad \mathrm{Tr}(Z^{[n]}) = \frac{(2n-L)d}{L},
	\end{equation}
	for $n\geq L/2$, the other half $n\leq L/2$ is obtained using the spin-inversion symmetry.

	Once the saturation value in each sector being obtained, we can easily obtain the saturation of the full operator by summing the sectors with the correct weights,
	\begin{equation}
		C_{L}(\infty) = \sum_{n=0}^{L} \frac{D_n}{2^{L}}C^{[n]}_{L}(\infty).
	\end{equation}
	Note that without using the symmetry, $C_{L}(\infty)$ would be exactly equal to $1$ for all $L$ due to the trace of $Z$ being $0$ in the full space: the symmetry prevents the
	$4$-point operator trace to vanish.
\end{appendices}

\end{document}